\documentclass[11pt,draft,onecolumn]{IEEEtran}

\usepackage[usenames]{color}

\usepackage{amsmath}
\usepackage{amssymb}
\usepackage{upref}
\usepackage{amsfonts}
\usepackage{graphicx}
\usepackage{pstricks-add}
\usepackage{fancyhdr,txfonts,pxfonts}
\usepackage{verbatim}
\usepackage{pgf,tikz}
\usepackage{mathrsfs}
\usetikzlibrary{arrows}
\usepackage[pagewise]{lineno}

\newcommand{\Fqn}{\mathbb{F}_q^n}
\newcommand{\mC}{\mathcal{C}}

\newcommand{\SU}{\mathrm{supp}}

\newcommand{\field}[1]{\mathbb{#1}}

\newcommand{\F}{\field{F}}

\newcommand{\N}{\field{N}}

\newcommand{\cG}{{\cal G}}

\newcommand{\cP}{{\cal P}}

\newcommand{\sP}{\cP}
\newcommand{\sG}{\cG}

\newcommand{\Gr}{\smash{{\sG\kern-1.5pt}_q\kern-0.5pt(n,k)}}
\newcommand{\Grtwo}{\smash{{\sG\kern-1.5pt}_2\kern-0.5pt(n,k)}}
\newcommand{\Gkone}{\smash{{\sG\kern-1.5pt}_q\kern-0.5pt(n,k_1)}}
\newcommand{\Gktwo}{\smash{{\sG\kern-1.5pt}_q\kern-0.5pt(n,k_2)}}
\newcommand{\Ps}{\smash{{\sP\kern-2.0pt}_q\kern-0.5pt(n)}}

\newtheorem{defn}{Definition}
\newtheorem{theorem}{Theorem}
\newtheorem{proposition}{Proposition}
\newtheorem{lemma}{Lemma}
\newtheorem{remark}{Remark}
\newtheorem{corollary}{Corollary}

\newtheorem{example}{Example}

\title{\huge Metrics Based on Finite Directed Graphs\\ and Coding Invariants}

\author{\textbf{Tuvi Etzion$^*$}, \textbf{Marcelo Firer$^+$}, \textbf{Roberto Assis Machado$^+$}\\
{\small $^*$Computer Science Department, Technion, Israel Institute of Technology, Haifa 32000, Israel}\\
{\small $^+$Department of Mathematics, Unicamp, Campinas, Brasil}\\
{\small {\it etzion@cs.technion.ac.il}, {\it mfirer@ime.unicamp.br}, {\it robertomachado@ime.unicamp.br}\vspace{-0.13ex}}
\thanks{This paper was presented in part at IEEE International Symposium
on Information Theory 2016 (ISIT 2016) and at IEEE Information Theory Workshop 2016 (ITW 2016).}
}

\begin{document}


\maketitle

\vspace{-1.9cm}


\begin{abstract}
Given a finite directed graph with $n$ vertices, we define a metric $d_G$ on $\F_q^n$, where $\F_q$
is the finite field with $q$ elements.
The weight of a word is defined as the number of vertices that can be reached by a directed
path starting at the support of the vector. Two canonical forms, which do not affect the metric,
are given to each graph. Based on these forms
we characterize each such metric. We further use these forms to prove that two graphs
with different canonical forms yield different metrics. Efficient algorithms to check if a
set of metric weights define a metric based on a graph are given. We provide
tight bounds on the number of metric weights required to reconstruct the metric.
Furthermore, we give a complete description of  the group of linear isometries of the graph metrics and a characterization of the graphs for which every linear code admits a $G$-canonical decomposition. Considering those graphs, we are able to derive an expression of the packing radius of linear codes in such metric spaces. Finally, given a directed graph which determines a hierarchical poset, we present sufficient and necessary conditions to ensure the validity of  the MacWilliams Identity and the MacWilliams Extension Property.

\end{abstract}

\section{Introduction}
\label{sec:introduction}

Let $\mathbb{F}_q^n$ be an $n$-dimensional vector space over a finite field $\F_q$ and let a code
be a linear subspace over $\mathbb{F}_q^n$. In order to gain some tools that may help (mainly in the decoding processes),
different metric structures are considered. The most common one, namely the Hamming metric, shares two well known properties:
\begin{enumerate}
\item It is determined by a weight, in the sense that $d(x,y)=w(x-y)$.
\item If we consider two error vectors $x=(x_1,\ldots,x_n)$ and $x^{\prime}=(x^{\prime}_1,\ldots,x^{\prime}_n)$
such that $x_i\neq 0$ if $x^{\prime}_i\neq 0$, then $w(x)\leq w(x^{\prime})$.
\end{enumerate}

Both properties are crucial in the context of coding theory. The first one means
that the distance $d(\cdot ,\cdot)$ is invariant by translations ($d(x+z,y+z)=d(x,y)$,
for all $x,y,z \in \F_q^n$), what  turns linear codes to be more
manageable in what concerns metric properties. To be more precise, as an example,
the well known Syndrome decoding algorithm performs a minimal distance decoding
procedure if and only if the metric is invariant by translations.
The second property says that the metric is suitable for error detection
and correction when considering reasonable channels, in the sense that making errors
in a family $I$ of bits can not be more probable then making errors in both the families $I$ and $J$.

Two large (and nearly disjoint) families of metrics on $\F_q^n$ which satisfy those
properties were introduced over the years: the combinatorial metric,
introduced by Gabidulin in 1971 \cite{Ga} and the poset metrics, introduced
in the context of Coding Theory by Brualdi et al. \cite{Brualdi} in 1995.

Since then, the poset metrics gained attention and has been extensively
studied in the literature  in the context of coding theory, exploring all its major aspects and invariants: MacWilliams Identity~ (\cite{Oh,KimOh}), perfect and MDS codes~\cite{Barg,MDS,Krotov}, duality, packing, and covering problems~\cite{Machado,Oliveira,Allan}.
Some generalizations of poset metrics were introduced over the years,
including its extension to Frobenius and finite principal ideal rings~\cite{barra},\cite{gre}
and the poset-block metrics \cite{FirerMunizPanek}.

In this work we introduce a new family of metrics: given a directed graph $G$, on $n$ vertices, we define a metric $d_G$  on $\F_q^n$. The graph based metrics generalize the poset metrics. It is closely related to the poset-block metrics (introduced in \cite{FirerMunizPanek}) but it is actually a new family of metrics. As we shall see in Section \ref{block}, the poset-block metrics may be obtained as a variation of the graph based metrics. In terms of coding theory, in some sense they work on different directions. If we consider, as an example, a situation where we have two blocks of information, blocks with different sizes (in bits), the poset-block values only the number of blocks where errors occur, but not the size of the blocks. The graph based metrics refine the poset-block metrics by considering the errors that compromise larger blocks to be more relevant than errors that damage the information on smaller blocks. 

Metrics defined by a graph may be useful to model some specific kind of channels and used to perform bitwise or message wise unequal error protection. The goal of this paper is to study such metrics in this context, and is organized as follows:

In Section~\ref{sec:basic} we present the basic definitions of directed graphs and
metrics defined by graphs. In Section~\ref{sec:canonical}, two canonical forms,
of directed graphs (for construction of the related metrics), are defined -- the expanded canonical form
and the reduced canonical form. We  prove that these two canonical forms are unique,
i.e. for a given directed graph there is a unique expanded canonical form and a unique
reduced canonical form. Moreover, we prove that two directed graphs determine the same metric if they have the same canonical forms. 
In Section~\ref{sec:computational} we consider several computational
questions related to the minimum number of metric weights which are required to
reconstruct the whole metric. These
questions have several variants which will be discussed in this section.
Bounds, some of which are tight are presented. From here on we move to questions closely related to coding theory. 
In Section~\ref{sec:coding}
we discuss the group of linear isometries of the graph metrics and the connection of
the work to coding theory. The structure of this group is a tool, which is used in Section \ref{candec} to determine an interesting decomposition of linear codes - the $G$-canonical decomposition - in the case when the  canonical reduced form of a graph is a hierarchical poset. Finally, in Section \ref{mac} we present some  conditions on a graph  that are sufficient for the validity of MacWilliams Identity and the MacWilliams Extension Property. 


\section{Basic Concepts}
\label{sec:basic}

A (simple finite) \emph{directed graph} $G(V,E)$ consists of a finite set
of \emph{vertices} $V=\{ v_1 ,\ldots , v_n\}$ and a set
of \emph{directed edges} $E$ (parallel edges are not allowed), where $e \in E$ is
an ordered pair $(u,v) \in V \times V$ with $u \neq v$. The vertex $u$ in this edge
is called the \emph{tail} of the edge and the vertex $v$ is called the \emph{head} of the edge.
A \emph{trail}  of \emph{length} $k$ is a sequence of edges $(u_0,u_1), (u_1,u_2),\ldots ,(u_{k-1},u_k)$
in which all edges are distinct. When $u_0 = u_k$, the trail is called a \emph{circuit}.
In case all the vertices $u_i$'s in a trail are distinct, except for the possibility
that $u_0 = u_k$, the trail is called  a \emph{simple directed path} of \emph{length}~$k$.
If $u_0=u_k$ then the path is called a \emph{directed cycle}. 

A \emph{complete graph} is a graph which contains all the possible
$n(n-1)$ edges. Such a graph is also called a \emph{clique}. 

If there is a trail from $u$ to $v$, we say that $u$ \emph{dominates} $v$,
and denote it by $u \rightarrow v$.  A set $X\subset V$ is called a \emph{closed set}
if $u\in X$ and $u$ dominates $v\in V$ implies that $v\in X$. The \emph{closure}
$\langle X\rangle_G $ of a set $X\subset V$ is the smallest closed subset containing $X$.
The set of all closed sets of $G$ is denoted by $\mathcal{I}(G)$. By abuse of notation,
if $X=\{ v\}$, we denote $\langle \{ v\}\rangle_G =\langle v\rangle_G$.

A directed graph $G(V,E)$ will be called \emph{L-weighted} (denoted by $G(V,E,L)$) if $G(V,E)$ is a directed graph and there
is a function $L:V \longrightarrow \N$, where $\N$ denotes the set of
natural numbers (positive integers). The value $L(v)$, $v \in V$, is called
the \emph{L-weight} of $v$. Clearly, if for each $v \in V$ we have that $L(v)=1$
then we can omit the L-weights of the vertices, and the L-weighted directed graph
is just a directed graph.

Each directed graph $G(V,E)$, where $V=\{ v_1 ,\ldots , v_n \}$,
defines a metric $d_G$ as follows:
The words of the space are all $n$-tuples over a given alphabet whose
size is at least two. Given $x=(x_1,x_2\ldots, x_n)\in\F_q^n$, the \emph{support}
of $x$ is the set of non-zero coordinates of $x$, i.e. $\mathrm{supp}(x)=\{i;x_i\neq 0\} $.
When considering an ordering (arbitrary but fixed) in the set $V=\{ v_1,v_2,\ldots, v_n\}$
of vertices of $G$, we have the $G$-\emph{support} of $x$, defined as
$\mathrm{supp}_G(x)=\{v_i\in V;x_i\neq 0\} $. For simplicity, we shall write
$\mathrm{supp}_G(x)=\mathrm{supp}(x)$. 
we can naturally identify $\mathrm{supp}(x)$ with  $\{i\in [n];x_i\neq 0\} $ and we shall do so if no confusion may arise.
The G-weight, $w_G(x)$, of a word $x$ is the number of distinct vertices in $G$
dominated by the vertices in the support of $x$:
\[
w_G(x)=|\langle \mathrm{supp}(x) \rangle_G|,
\]
where $|A|$ denotes the cardinality of the set $A$.
The $G$-\emph{distance} between two
words $x, y \in \F_q^n$ is defined by $d_G(x,y)=w_G(y-x)$. 
Since the actual values in the nonzero entries of $x$ and $y$ do not play any role in the determination of the G-weights,
we will assume, without loss of generality, that all the G-weights are given only for binary words.

\begin{theorem}
If $G(V,E)$ is a directed graph, then $d_G$ is a metric. 
\end{theorem}

\begin{example}
	The \emph{Hamming weight} $w_H$ is a particular case of a $G$-weight where $E=\emptyset$. On the opposite side, if $G$ is a complete graph, then  $\langle A \rangle_G=V$ for every $A\ne\emptyset$ and hence, $w_G(x)=n$, for each $x\neq 0$.
\end{example}
\section{Canonical Forms}
\label{sec:canonical}

In this section two canonical forms will be introduced for any given finite
directed graph $G(V,E)$, one reduced canonical form and the other expanded canonical form.
Each canonical form can be served to characterize a set of graphs which form
the same G-metric. These canonical forms are unique and
will lead to one of the main results of this paper,
that two different graphs whose expanded or reduced canonical forms are different
yield two different metrics. The canonical forms can be defined through
a set of edges called shortcuts.
An edge $(u,v)\in E$ is called a \emph{shortcut} if there exists
a simple directed path from $u$ to $v$ which contains at least two edges.
Adding or removing shortcuts to the graph
do not affect the metric as the following lemma, whose proof is obvious, states.

\begin{lemma}
\label{lem:shortcut}
If $G(V,E)$ and $G'(V,E')$ are two directed graphs which differ in exactly one edge $e$,
and $e$ is a shortcut, then $d_G=d_{G'}$.
\end{lemma}

We remark that shortcuts must be removed one at a time and not simultaneously,
in order to avoid the situation which we find in the next example:

\begin{example}\label{examp:atalhofalso}
Let $G(V,E)$ be a graph with $V=\{u,v,w\}$ and $E=\{ (u,v),(u,w),(v,w),(w,v) \}$. Then $(u,v)$ and $(u,w)$ are shortcuts.
By removing any of those shortcuts or all of them we get the sets of edges
$E_v=\{ (u,w),(v,w),(w,v) \}$, $E_w=\{ (u,v),(v,w),(w,v) \}$ and
$E_{v,w}=\{ (v,w),(w,v) \}$, to which corresponds the graphs $G_v$, $G_w$ and $G_{v,w}$, respectively. One can directly check that
\begin{align*}
\langle u\rangle_{G} &=\langle u\rangle_{G_v}=\langle u\rangle_{G_w}=\{u,v,w\} \\
\langle u\rangle_{G_{v,w}}&=\{ u\}.
\end{align*}
\end{example}
The \emph{expanded canonical form} of a directed graph $G(V,E)$ is the graph $G'(V,E')$ obtained by adding edges, one by one, in a way that each edge added to the graph is a shortcut. Hence, the expanded canonical graph $G'(V,E')$ is a graph for which $E \subset E'$, $E' \setminus E$ contains only shortcuts and any possible edge to add in $E'$ is not a shortcut.

As an example, the expanded canonical form of a graph is a complete graph if and only if there is a circuit containing all the vertices. The expanded canonical form of a graph $G(V,E)$ has all the possible shortcuts. Some of these induce complete subgraphs of the expanded canonical form $G'(V,E')$. To figure out the maximal cliques in the expanded canonical form we state the next elementary lemmas, which follow straightforward from the definition of $G'(V,E')$.

\begin{lemma}
Let $G(V,E)$ be a directed graph and $G'(V,E')$ its expanded canonical form. Let $V'\subset V$ be such that  $u$ dominates $v$  for any $u,v\in V'$. Then $G'(V,E')$ induces a clique on $V'$.
\end{lemma}

The clique induced on $V'$, by the expanded canonical form $G'(V,E')$, is called \emph{maximal} if there is no $v\in V\setminus V'$ such that $G'(V,E')$ induces a clique on $V'\cup \{v\}$. It is clear that if $V'$ and $V''$ are maximal cliques, then either $V'=V''$ or $V' \cap V'' = \varnothing$. Finally, we have that the expanded canonical form is unique.
\begin{lemma}
\label{lem:unique_expanded}
Any directed graph has a unique expanded canonical form.
In other words, the expanded canonical form is well-defined.
\end{lemma}
\begin{theorem}
\label{thm:same_metric}
Two different graphs with the same expanded canonical form yield the same metric.
\end{theorem}
\begin{IEEEproof}
Let $G(V,E)$ be a graph and $G'(V,E')$ its canonical expanded form. Since we can move from the expanded graph $G'$ to the original graph $G$ by removing one by one the shortcuts in $E' \setminus E$, Lemma \ref{lem:shortcut} ensures that $d_G=d_{G'}$.
\end{IEEEproof}

Next, we want to define a second canonical form for a graph which defines a metric. Succinctly, the \emph{reduced canonical form} is an $L$-weighted acyclic graph  $G'(V',E',L')$ obtained from a graph $G(V,E)$ by contracting each strongly-connected component to a vertex, and the weight of this vertex is the number of vertices in the strongly-connected component, i.e., assuming that $G(V,E)$ is in the expanded canonical form (Lemma \ref{lem:unique_expanded}, we obtain $G'(V',E',L')$ in the following way:

\begin{enumerate}
\item Each maximal clique in $G(V,E)$ is associated to a vertex $u$ in $V'$. We denote by $\pi:V\rightarrow V'$ the map that associates to a vertex of $V$ the vertex of $V'$ that represents its clique.

\item The weight $L':V'\longrightarrow \N$ is defined as $L'(u)=|\pi^{-1}(u)|$.

\item The projection $\pi$ also defines the structure of edges: we start defining a set of vertices $E^{\ast}$ where  $(u_1,u_2)\in E^{\ast}$ if, and only if, there are $v_1\in\pi^{-1}(u_1)$ and $v_2\in\pi^{-1}(u_2)$ such that $v_2\in \langle v_1\rangle_G$. We remark that $G^{\ast}=(V',E^{\ast})$ has no circuits, so that it actually determines a partial order on $V'$: $u_1\preceq u_2$ if $u_1\in\langle u_2 \rangle_{G^{\ast}}$. We denote this poset by $P_G=(V',\preceq)$. It follows that, on $G^{\ast}$, the closure of a set $A$ is just the ideal generated by $A$ in the usual meaning in the context of posets.

\item Finally, we obtain $E'$ by removing \emph{all} shortcuts from $E^{\ast}$ and get the \emph{canonical reduced form} $G'(V',E',L')$.
\end{enumerate}

We remark the following:

\begin{itemize}
\item The \emph{reduced closure} of a subset $A\subset V$ is the $G'$ closure under the projection $\pi$, i.e., $\langle A\rangle_{GL':V'\longrightarrow \N}:=\langle \pi (A)\rangle_{G'}$. By doing so, given $x\in\Fqn$ we can determine its $G$-weight by considering the $L$-weighted reduced form of $G$:
\[
w_G(x)=\sum_{u\in \langle \mathrm{supp} (x)\rangle_{G'}}L(u).
\]
 We remark that, if we define a constant weight $L_1(u)=1$, for all $u\in V'$, we get the poset-block metric as defined in \cite{FirerMunizPanek}. \label{block} 
\item There may be many shortcuts with tail $u$ and  the heads $\{ u_1,u_2,\ldots, u_r\}$
contained in a single clique. In the reduced form, all these edges are replaced
by a single edge with tail $u$ and head in the vertex that replaces the clique.
\end{itemize}

	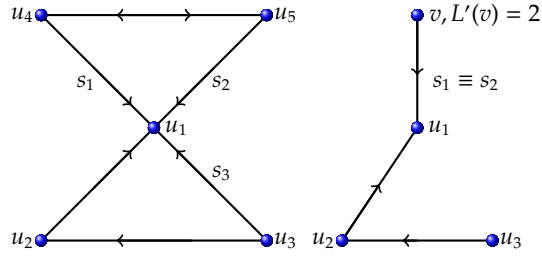
\begin{figure}[h]
		\centering
		\begin{tikzpicture}
		
		\draw[thick][-] (-1.5,1.5) -- (1.5,-1.5);
		\draw[thick][-] (-1.5,-1.5) -- (1.5,1.5);
		\draw[thick][-] (-1.5,1.5) to (1.5,1.5);
		\draw[thick][-] (-1.5,-1.5) to (1.5,-1.5);
		
		\draw[thick][->] (-0.8,0.8) -- (-0.3,0.3);
		\draw[thick][->] (0.8,-0.8) -- (0.3,-0.3);
		
		\draw[thick][->] (0.8,0.8) -- (0.3,0.3);
		\draw[thick][->] (-0.8,-0.8) -- (-0.3,-0.3);
		
		\draw[thick][<->] (-0.5,1.5) -- (0.5,1.5);
		\draw[thick][<-] (-0.5,-1.5) -- (0.5,-1.5);
		
		\draw[thick][-] (3.5,1.5) -- (3.5,0);
		\draw[thick][<-] (3.5,0.7) -- (3.5,1.5);
		
		\draw[thick][->] (2.5,-1.5) -- (3,-0.75);
		
		\draw[thick][-] (4.5,-1.5) -- (2.5,-1.5);
		\draw[thick][->] (3.5,-1.5) -- (3.3,-1.5);
		\draw[thick] (2.5,-1.5) -- (3.5,0);
		
		
		\draw plot[only marks, mark=ball, mark size={2.3}] coordinates{(0,0)(-1.5,1.5)(1.5,1.5)(-1.5,-1.5)(1.5,-1.5)(3.5,1.5)(2.5,-1.5)(4.5,-1.5) (3.5,0)};
		
		\draw(0.3,0)  node {\footnotesize{$u_1$}};
		\draw(-1.75,-1.5)  node {\footnotesize{$u_2$}};
		\draw(1.75,-1.5)  node {\footnotesize{$u_3$}};
		\draw(-1.75,1.5)  node {\footnotesize{$u_4$}};
		\draw(1.75,1.5)  node {\footnotesize{$u_5$}};
		\draw(0.9,0.6)  node {\footnotesize{$s_2$}};
		\draw(-0.9,0.6)  node {\footnotesize{$s_1$}};
		\draw(0.9,-0.6)  node {\footnotesize{$s_3$}};
		\draw(3.8,0)  node {\footnotesize{$u_1$}};
		\draw(2.25,-1.5)  node {\footnotesize{$u_2$}};
		\draw(4.75,-1.5)  node {\footnotesize{$u_3$}};
		\draw(4.4,1.5)  node {\footnotesize{$v, L^{\prime}(v)=2$}};
		\draw (4.15,0.6) node {\footnotesize{$s_1\equiv s_2$}};
		\end{tikzpicture}
		\caption{On the left a directed graph with 5 vertices and one single clique $\{u_4,u_5\}$. On the right, its reduced canonical form. The clique is replaced by a vertex $v$ with $L^{\prime}(v)=2$. Whenever we removed the shortcut $s_1$ or $s_2$ is immaterial, since they are replaced by a single directed edge $(u_1,v)$.}
\end{figure}

\begin{proposition}
\label{lem:unique_reduced}
Any directed graph has a unique reduced canonical form, i.e., it is well-defined.
\end{proposition}
\begin{IEEEproof}
The structure of maximal cliques in $G$ is unique and so the graph $G^{\ast}$
is uniquely determined by $G$. Moreover, the $L'$-weight is determined exclusively
by the projection $\pi:V\longrightarrow V'$, hence it is also uniquely determined.
$G^{\ast}$ is a poset and $G'$ may be viewed as its Hasse diagram. It is well
known that the Hasse diagram (considered as a directed graph) of a poset is uniquely
determined by the poset and it follows that the reduced canonical form is well-defined.
\end{IEEEproof}

Proposition~\ref{lem:unique_reduced} leads to the two main results of this section.
Before we state and prove them, we need a simple and useful lemma:

\begin{lemma}
\label{lemma:uv}Let $G_{1}\left(  V,E_{1}\right)  $ and
$G_{2}\left(  V,E_{2}\right)  $ be two graphs such that $d_{G_1} =d_{G_{2}}$. Then, given $u,v\in V$, $v\in\left\langle
u\right\rangle _{G_{1}}$ if and only if $v\in\left\langle u\right\rangle
_{G_{2}}.$
\end{lemma}
\begin{IEEEproof}
Suppose that $v\in\left\langle u\right\rangle _{G_{1}}$ and $v\notin
\left\langle u\right\rangle _{G_{2}}$. This would imply that $\left\langle
u\right\rangle _{G_{1}}=\left\langle \left\{  u,v \right\}  \right\rangle
_{G_{1}}$ while $\left\langle u\right\rangle _{G_{2}}\varsubsetneqq
\left\langle \left\{  v,u\right\}  \right\rangle _{G_{2}}$ and hence
a word $x\in\mathbb{F}_{q}^{n}$ such that $\mathrm{supp}\left(
x\right)  =\left\{  u,v\right\}  $ will have different weights: $w_{G_{1}%
}\left(  x\right)  <w_{G_{2}}\left(  x\right)$, contradicting
the assumption that $d_{G_{1}}  =d_{G_{2}}$.
\end{IEEEproof}

\begin{theorem}
\label{can_exp}
If two directed graphs $G_1(V,E_1)$ and $G_2(V,E_2)$ induces metrics
such that $d_{G_1}=d_{G_2}$ then the expanded canonical forms of $G_1$
and $G_2$ are equal, and the same is true for their reduced canonical forms.
\end{theorem}
\begin{IEEEproof}
The proof is done by induction on $n=\left\vert V\right\vert $. The proof of the base $n=1$
is trivial. Let $0\neq e_{v}\in\mathbb{F}%
_{q}^{n}$ be a word with $\mathrm{supp}\left(  e_{v}\right)  =\left\{  v\right\}  $
and with minimal $w_{G_{1}}$ and $w_{G_{2}}$ weight. If $w_{G_{1}}\left(
e_v\right)  =w_{G_{2}}\left(  e_v\right)  =n$, then both $G_{1}$ and $G_{2}$
contain a cycle of length $n$ hence they expanded canonical form is a complete
graph with $n$ vertices and the reduced canonical form is the graph with a
unique vertex $u$ and $L_{1}\left(  u\right)  =L_{2}\left(  u\right)  =n$.

We assume that $w_{G_{i}}\left( e_v\right)  <n$. We
consider the projections $u_{1}=\pi_{1}\left(  v\right)  $ and $u_{2}=\pi
_{2}\left(  v\right)  $ where $\pi_{i}:G_{i}\rightarrow G_{i}^{\prime}$ \ is
the projection onto the canonical reduced form, for $i=1,2$. The minimality of
$w_{G_{i}}\left(  v\right)  $ ensures that $u_{i}$ is a minimal element in the
poset $G_{i}^{\prime}$ and $w_{G_{i}}\left(  v\right)  =L_{i}\left(
u_{i}\right)  =\left\vert \pi_{i}^{-1}\left(  u_{i}\right)  \right\vert $, for
$i=1,2$. It follows that the maximal cycle containing $v$ in $G_{1}$ and
$G_{2}$ ($\pi_{1}^{-1}\left(  u_{1}\right)  $ and $\pi_{2}^{-1}\left(
u_{2}\right)  $, respectively) have the same cardinality and if $\left(
v,w\right)  \in E_{i}$, then $w\in\pi_{i}^{-1}\left(  u_{i}\right)  $. We
consider now the graphs $G_{i}^{\ast}$ obtained from $G_{i}$ by adding all the
possible edges to the maximal cycle $\pi_{i}^{-1}\left(  u_{i}\right)  $. It
is clear that $d_{G_{i}}=d_{G_{i}^{\prime}}$, for
$i=1,2$. Let us consider the graph $\Gamma_{i}$ obtained from $G_{i}^{\prime}$
by removing the vertex $v$ and all the edges of $G_{i}^{\prime}$ that have $v$
either as a tail or as a head. It follows that
\[
\left\vert \left\langle w\right\rangle _{\Gamma_{i}}\right\vert =\left\{
\begin{array}
[c]{c}%
\left\vert \left\langle w\right\rangle _{G_{i}}\right\vert \text{ if }%
v\not\in\left\langle w\right\rangle _{G_{i}}\\
\left\vert \left\langle w\right\rangle _{\Gamma_{i}}\right\vert -1\text{ if
}v\in\left\langle w\right\rangle _{G_{i}}%
\end{array}
\right.  \text{.}%
\]
Lemma~\ref{lemma:uv} ensures that $v\in\left\langle w\right\rangle _{G_{1}}$
if and only if $v\in\left\langle w\right\rangle _{G_{2}}$ and it follows
that $d_{\Gamma_{1}}=d_{\Gamma_{2}}$. By the
induction hypothesis, we may assume that the reduced and the extended
canonical forms of $\Gamma_{1}$ and $\Gamma_{2}$ are the same. Let $\Gamma
_{1}^{\prime}=\Gamma_{2}^{\prime}=\left(  V^{\prime},E^{\prime},L^{\prime
}\right)  $ be the reduced form of both $\Gamma_{1}$ and $\Gamma_{2}$. If
$w_{G_{i}}\left(  v\right)  >1$ then the canonical reduced form of $G_{i}$ is
the $L$-weighted graph  $\left(  V^{\prime},E^{\prime},L\right)  $, wich
differ from $\Gamma_{i}^{\prime}$ only in the $L$-weight:
\[
L\left(  u\right)  =\left\{
\begin{array}
[c]{c}%
L^{\prime}\left(  u\right)  \text{ if }u\in\pi_{i}\left(  v\right)  \\
L^{\prime}\left(  u\right)  +1\text{ if }u=\pi_{i}\left(  v\right)
\end{array}
\right.  \text{.}%
\]
If $w_{G_{i}}\left(  v\right)  =1$, we have that canonical reduced form of
$G_{i}$ is the $L$-weighted graph $G_{i}^{\prime\prime}=\left(  V_{i}%
^{\prime\prime},E_{i}^{\prime\prime},L_{i}^{\prime\prime}\right)  $ where
\begin{align*}
V_{i}^{\prime\prime}  & =V^{\prime}\cup\left\{  u_{i}\right\}  ,\\
E_{i}^{\prime\prime}  & =E^{\prime}\cup\left\{  \left(  \pi_{i}\left(
w\right)  ,u_{i}\right)  ;v\in\left\langle w\right\rangle _{G_{i}}\right\}
,\\
L_{i}^{\prime\prime}\left(  u\right)    & =\left\{
\begin{array}
[c]{c}%
L^{\prime}\left(  u\right)  \text{ for }u\neq u_{i}\\
1\text{ for }u=u_{i}%
\end{array}
\right.  \text{,}%
\end{align*}
recalling that $u_{i}=\pi_{i}\left(  v\right)  $. Again, Lemma \ref{lemma:uv}
ensures $v\in\left\langle w\right\rangle _{G_{1}}$ if, and only if,
$v\in\left\langle w\right\rangle _{G_{2}}$ and it follows that, up to renaming
$u=u_{1}=u_{2}\,\ $\ we have that $E_{1}^{\prime\prime}=E_{2}^{\prime\prime}$
hence $G_{1}^{\prime\prime}=G_{2}^{\prime\prime}$.

For the extended canonical form we proceed in exactly the same way, making the
induction step by excluding a vertex $v\in V$ such that $\left\vert
\left\langle v\right\rangle _{G_{1}}\right\vert =\left\vert \left\langle
v\right\rangle _{G2}\right\vert $ is minimal.
\end{IEEEproof}

\begin{corollary}
Two graph metrics  are isomorphic if and only if their related graphs
in expanded or reduced canonical forms are isomorphic.
\end{corollary}
\begin{IEEEproof} It follows from Theorem \ref{can_exp}, Theorem \ref{thm:same_metric} and from the fact that $$w_G(x)=\sum_{u\in \langle \mathrm{supp} (x)\rangle_{G'}}L(u).\vspace{-22pt} $$
 \end{IEEEproof}


\section{Metric Reconstruction}
\label{sec:computational}

In this section we will consider the minimum number of G-weights which
are required to reconstruct the whole metric. This question has at least three variants.

\begin{enumerate}
\item Let $S(n)$ be a set of words of length $n$ such that for any given metric based on a directed graph with $n$ vertices, the $G$-weights of the words in $S(n)$ are sufficient to recover the whole metric. Let $D(n)$ be the minimum size of $S(n)$. What is the value of $D(n)$?

\item Let $d_G$ be a metric based on a directed graph with $n$ vertices. Let $M_G(n)$ be minimum number of G-weights required to recover the whole metric. Let $G$ be the expanded canonical graph for which this number is the smallest among all graphs, and let $M^{min}(n)$ be the value of $M_G(n)$ for this graph. Let $G$ be the expanded canonical graph for which this number is the largest among all graphs, and let $M^{max}(n)$ be the value of $M_G(n)$ for this graph. What is the value of $M^{min}(n)$? What is the value of $M^{max}(n)$?

\item Given a metric $d_G$, we can ask queries, each query on the G-weight value of a specific word, in a sequence, where each query is based on the values given in previous answers that we got. Let $Q(n)$ be the minimum number of queries which are required to recover any given such metric. What is the value of~$Q(n)$?

\end{enumerate}

We note on the difference between the second and the third variants.
For the second variant we assume that for the given metric we receive the smallest number of metric weights to recover all the G-weights. In the third variant we have to find a general strategy to ask queries in a way that the total number of queries will be small.

The following lemma is readily verified.
\begin{lemma}
\label{lem:4_values}
$M^{min}(n) \leq M^{max} (n) \leq Q(n) \leq D(n)$.
\end{lemma}

We start with the fundamental question of constructing the graph from the
G-weights of the words in the space. Given a set of all G-weights we would like to
find the related graph, i.e. its reduced canonical form or its expanded canonical form.
To this end we first need the following lemma which is verified from the definitions.

\begin{lemma}
\label{lem:dominate}
For a given graph metric $d_G$, of a graph $G(V,E)$, $w_G(e_i + e_j) = w_G(e_i)$, $i \neq j$, if and only if
there exists a path from $v_i$ to $v_j$, i.e. $v_i$ dominates $v_j$.
\end{lemma}


\begin{theorem}
The G-weights of all words with Hamming weights one or two are sufficient
to determine the expanded canonical form of the related graph.
\end{theorem}
\begin{IEEEproof}
By Lemma~\ref{lem:dominate}, we can determine for each pair of vertices
$\{ v_i , v_j \}$, $i \neq j$, if the edges $v_i \rightarrow v_j$ and
$v_j \rightarrow v_i$ exist or not, by observing G-weights only of
words with Hamming weight one or two.
\end{IEEEproof}

\begin{corollary}
\label{cor:two_weights}
If the graph $G$ with the metric $d_G$
has $n$ vertices, then the $n + \binom{n}{2}$ G-weights of words
with Hamming weights one or two are sufficient to determine the whole metric.
\end{corollary}

\begin{corollary}
\label{cor:weakDn}
$D(n) \leq n + \binom{n}{2}$.
\end{corollary}

On the other hand it is not difficult to see that there are metrics in which
all the weights, except for one, are not sufficient to determine the whole metric. Consider the following
two graphs $G_1(V,E_1)$ and $G_2(V,E_2)$, where $V=\{ v_1 , \ldots , v_n \}$. Assume that
$G_1$ is the complete graph, while in $G_2$, $v_1 , \ldots , v_{n-1}$ form a clique
on $n-1$ vertices, and there is an edge $v_i \rightarrow v_n$, for each
$i$, $1 \leq i \leq n-1$. The G-weights of all words, except one word $e_n$, in these two metrics are
the same ($w_{G_1}(x)=w_{G_2}(x)=n$ for each word $x \in \F_q^n$, except for $w(G_1)_{e_n} =n$ and $w_{G_2}(e_n)=1$).

\begin{corollary}
\label{cor:weight_one}
If $d_G$ is a metric based on a graph $G$,
then each one of the G-weights of words with Hamming weight one
might be required to determine the whole metric, i.e. $e_i \in S(n)$
for each $1 \leq i \leq n$.
\end{corollary}

Corollary~\ref{cor:weakDn} can be slightly improved as follows.

\begin{theorem}
\label{thm:strongDn}
$D(n) \leq \lceil \frac{n}{2} \rceil + \binom{n}{2}$.
\end{theorem}

\begin{IEEEproof}
Let $V = \{ v_1 ,\ldots , v_n \}$ be the set of vertices of $G$. We claim that
the G-weights of all the words of Hamming weight one and the
G-weights of all the words with Hamming weights two, excluding $w_G(e_1+e_2)$,
$w_G(e_3+e_4)$, $w_G(e_5+e_6)$,..., are sufficient to recover the G-weights
of the whole metric. To prove this we only have to show that we can find the
weights of $w_G(e_1+e_2)$, $w_G(e_3+e_4)$, $w_G(e_5+e_6)$,..., and apply
Corollary~\ref{cor:two_weights}. By Lemma~\ref{lem:dominate}, we have that
by observing the set $\{ w_G(e_1+e_i) ~:~ 3 \leq i \leq n \}$ we can find
how many (and which) vertices (excluding $v_2$) are dominated by $v_1$. Let $\delta_1$ be
the number of vertices that $v_1$ dominates (including $v_1$ and excluding
$v_2$). If $w(e_1)=\delta_1$ then clearly $v_1$ does not dominate $v_2$
and if $w_G(e_1)=\delta_1 +1$ then clearly $v_1$ dominates $v_2$.
Similarly, let $\delta_2$ be
the number of vertices that $v_2$ dominate (including $v_2$ and excluding
$v_1$) and similarly we can find if $v_2$ dominates $v_1$ or not. Now, if $v_1$ dominates $v_2$
then $w_G(e_1 + e_2)=w_G(e_1)$ and if $v_2$ dominates $v_1$ then $w_G(e_1+e_2)=w_G(e_2)$.
Finally, if $v_1$ does not dominate $v_2$ and $v_2$ does not dominate $v_1$
then $w_G(e_1+e_2) =\delta_1 + \delta_2 -t$, where $t$ is the number of vertices
dominated by both $v_1$ and $v_2$ (and $t$ can be found by observing the
vertices dominated by $v_1$ and observing the vertices dominated by $v_2$).
Similarly, we can find $w_G(e_3+e_4)$ and so
on. To conclude, we apply Corollary~\ref{cor:two_weights} to complete the proof.
\end{IEEEproof}

\begin{theorem}
Given any set with less than
$\lceil \frac{n}{2} \rceil + \binom{n}{2}$ G-weights, there exists
a directed graph $G$ with $n$ vertices whose graph metric cannot be reconstructed
only from these G-weights.
\end{theorem}
\begin{IEEEproof}
Consider the following two directed graphs $G_1(V,E_1)$ and $G_2(V,E_2)$,
where $V= \{ v_1 ,\ldots , v_n \}$.
$G_1$ has a clique on the set of vertices
$\{ v_4 ,\ldots , v_n \}$; for each $i$, $4 \leq i \leq n$,
$G_1$ has the edge $v_i \rightarrow v_j$, $1 \leq j \leq 3$, and it also has the edge $v_1 \rightarrow v_2$.
$G_2$ has a clique on the set of vertices
$\{ v_4 ,\ldots , v_n \}$; for each $i$, $4 \leq i \leq n$,
$G_2$ has the edge $v_i \rightarrow v_j$, $1 \leq j \leq 3$, and it also has the edge $v_1 \rightarrow v_3$.
This means that the two graphs differ only in the edges $v_1 \rightarrow v_2$ and $v_1 \rightarrow v_3$.
The related metrics differ only in two G-weights.
In $G_1$, $w_{G_1}(e_1+e_2)=2$ and $w_{G_1}(e_1+e_3)=3$, while
in $G_2$, $w_{G_2}(e_1+e_2)=3$ and $w_{G_2}(e_1+e_3)=2$. Hence, if these two
G-weights are not given we won't be able to distinguish between the two metrics.
Therefore, from the G-weight of words with Hamming weight two only a set of G-weights
which correspond to vertex-disjoint edges can be omitted.
Thus, from all the G-weights of words with Hamming weight two we are required
to have at least $\binom{n}{2} - \lfloor \frac{n}{2} \rfloor$ words. By Corollary~\ref{cor:weight_one},
the G-weights of all the words with Hamming weight one are also
required and the lemma follows.
\end{IEEEproof}

\begin{corollary}
\label{cor:lastDn}
$D(n) \geq \lceil \frac{n}{2} \rceil + \binom{n}{2}$.
\end{corollary}

Now, by Theorem~\ref{thm:strongDn} and Corollary~\ref{cor:lastDn} we have that

\begin{corollary}
$D(n) = \lceil \frac{n}{2} \rceil + \binom{n}{2}$.
\end{corollary}

The values of $M^{min}(n)$ and $M^{max}(n)$ given in the next two theorems.
\begin{theorem}
$~$
$M^{min}(n)=n$.
\end{theorem}
\begin{IEEEproof}
Given the Hamming metric which is represented by a graph $G$
with $n$ isolated vertices, it is trivial to see that
the set of weights $\{ w_G (e_i ) ~:~ i \in [n] \}$ is sufficient
to determine the metric. Hence, $M^{min}(n) \leq n$.

$M^{min}(n) \geq n$ from information theory arguments.
\end{IEEEproof}

\begin{lemma}
\label{lem:lowerM}
$M^{max}(n) \geq 2n-4$.
\end{lemma}
\begin{IEEEproof}
If $n=1$ then no queries are required and if $n=2$ then two queries are
necessary. For $n=3$ the graph with exactly one edge requires four queries.

Now, we have to present a graph with $n \geq 4$ vertices that requires at least
$2n-4$ queries to reveal all its edges. The graph will consists of at least four
vertices. One vertex $v_{max}$ dominates all the other vertices.
One vertex $v_{min}$ which does not dominate any other vertex.
Any other vertex of the $n-2$ vertices dominates $v_{min}$. We must know $w_G(v_{max})$ to
reveal it dominates all the other vertices. For any vertex $v$ which dominates
$v_{min}$ we must know $w_G(v)$, but this is not enough. For example, assume
that $w_G (v_1)=w_G(v_2)=2$. If we have no other information on $v_1$ and $v_2$ they
can either dominate $v_{min}$ or be together in a clique of size two. Hence, we need
two queries for each one of the vertices that dominates $v_{min}$ with a possible exception
of one. Revealing all this information implies that $w_G (v_{min})=1$, so there is no need
to have a query on $v_{min}$. Thus, the total number of required queries is $2n-4$.
\end{IEEEproof}

\begin{lemma}
\label{lem:upperM}
$M^{max}(n) \leq 2n-1$.
\end{lemma}
\begin{IEEEproof}
For a given vertex $u$, that dominates the vertices
$v_1,v_2,\ldots , v_r$, the two G-weights $w_G (u)$
and $w_G (u,v_1,v_2,\ldots,v_r )$, which are equal to $r+1$ are enough
to know all the edges from $u$ to the other vertices of the graph.
This proves that $M^{max}(n) \leq 2n-2$. Hence, we only have to show that
we can omit one G-weight. For any vertex $u$ which does not dominate any
other vertex, the G-weight $w_G(v)$ is sufficient and if there ais one such vertex
then the claim of the lemma is proved. If there is no such vertex, then the
vertex of minimal G-weight is in a clique of size $r \geq 2$, which consists of the vertices
$v_1,v_2,\ldots , v_r$. The $r+1$ G-weights $w_G (v_1),w_G(v_2),\ldots,w_G(v_r)$, and
$w_G(v_1,v_2,\ldots,v_r)$ are sufficient to find this clique and hence $2n-1$ G-weights
are sufficient.

Thus, $M^{max}(n) \leq 2n-1$.
\end{IEEEproof}

The gap between the lower bound of Lemma~\ref{lem:lowerM} and the upper bound
of Lemma~\ref{lem:upperM} can be reduced, but we omit it as the proof is too tedious
and we leave it to the reader.
For the value of $Q(n)$, we managed only slightly to bridge on the gap
between the upper and lower bounds derived from Lemma~\ref{lem:4_values}. Hence, the intriguing
problem to determine the value of $Q(n)$ is left as a research problem for the reader.

After the expanded canonical form graph was constructed based on the G-weights of the
words with Hamming weights one and two, we would like to know whether
the given G-weights of the other words are consistent with the G-weights
of the words with Hamming weights
one or two. The most simple way is to consider
each $r$-subsets of vertices and use a search algorithm for all the vertices
dominated by this $r$-subset of vertices.
We leave other variants to the interested reader and for future research.hs).
Some of the variants are
related to the well-known question concerning graph
isomorphism, until recently the only fundamental computational problem
whose answer is completely unknown~\cite{Tor04}
(before the most recent result of L\'{a}szl\'{o} Babai~\cite{Bab16}).

\section{Isometries and Coding}
\label{sec:coding}


As seen in Section \ref{sec:canonical}, the reduced canonical form $G'(V',E',L')$ of $G$ may be considered as a poset $P_G=(V',\preceq)$, and we say that  $v\in X$ is \emph{maximal} in $X\subset V$ if
$\pi (v)$ is maximal in $\pi (X)\subset V'$, considered as a subposet of $P_G$. We denote by $\emph{Max}_G(X)$ the set of maximal elements of $X$ and  $\emph{Max}_G (x):=\emph{Max}_G (\mathrm{supp}(x))$. The  \emph{cleared out form} of $x\in\Fqn$ is the  vector $\tilde{x}=(\tilde{x}_1, \ldots ,\tilde{x}_n) $ such that $\tilde{x}_i=x_i$ if $i\in \emph{Max}_G(x)$ and $\tilde{x}_i=0$ otherwise. 
 A subset $\widetilde{X}\subset X\subset V$ is said to be a \emph{minimal set of generators}
(MSG) of $X$ if $\left\langle \widetilde{X}%
\right\rangle _{G}= \left\langle X\right\rangle _{G}$ 
and
$\widetilde{X}$ is minimal with this property.  A MSG is not unique, but it is easy to see that   $\widetilde{X}\subset \emph{Max}_G(X)$.
A \emph{minimal set of generators}
(MSG) of $X\subset V$ is a   subset $\widetilde{X}\subset X$ such that $\left\langle \widetilde{X}%
\right\rangle _{G}= \left\langle X\right\rangle _{G}$ 
and
$\widetilde{X}$ is minimal with this property.  A MSG is not unique, but it is easy to see that   $\widetilde{X}\subset \emph{Max}_G(X)$.


Given a directed graph $G$ we denote by $\widetilde{G}$ its expanded canonical form of $G$,  $Aut\left( \widetilde{G} \right)  $ the group of automorphisms of $\widetilde{G}$  and $GL\left(  n,G\right)  _{q}$ the group of linear
	isometries of $\left(  \mathbb{F}_{q}^{n},d_G\right)  $, i.e.,
\begin{align*}
GL\left(  n,G\right)_{q}=\{   T  : \mathbb{F}_{q}^{n}\rightarrow
\mathbb{F}_{q}^{n} ; \ T \text{ is linear and }
d_{G}\left(  x,y\right)  =d_{G}\left(T  \left( x\right) ,T\left( y\right)  \right), \forall x,y\in \mathbb{F}_{q}^{n} \}.
\end{align*}
We first remark is that  $GL\left(  n,G\right)_{q} =  GL\left(   n,\widetilde{G}\right)_{q}$, so  an automorphism $\phi\in Aut\left(
	\widetilde{G}\right)$ induces an isometry $T_{\phi}\in GL\left(  n,G\right)_{q}$,
	acting on $\mathbb{F}_{q}^{n}$ by permutation of the coordinates: $T_{\phi
	}\left(  (  x_{1},x_{2},\ldots,x_{n})  \right)  =(
	x_{\phi\left(  1\right)  },x_{\phi\left(  2\right)  },\ldots,x_{\phi\left(
		n\right)  })$.



We say that a linear map $T:\mathbb{F}_{q}^{n}\rightarrow\mathbb{F}%
_{q}^{n}$ \emph{respects domination} if $T\left(  e_{i}\right)  =\sum_{j=1}^{n}\alpha_{ij}e_{j}$ satisfies the conditions: 
\emph{(i)} $\alpha
_{ii}\neq0$ for every $i \leq n$; \emph{(ii) }$\alpha_{ij}\neq0$
implies that $v_{j}\in\left\langle v_{i}\right\rangle _{G}$.
We denote  $N\left(  G\right)$ as the set of all linear maps respecting domination.

\begin{lemma}
$Aut\left(  \widetilde{G}\right)  $ and $N\left(  G\right)  $ are subgroups of $GL\left(
n,G\right)  _{q}$.
\end{lemma}

\begin{IEEEproof}
It is clear that if $T_{\phi}\in Aut\left(  \widetilde{G}\right)$, then  $T_{\phi}\in GL\left(
n,G\right)  _{q}$. To conclude that  $Aut\left(  \widetilde{G}\right)$ is a subgroup, note that $T_{\phi}T_{\psi}^{-1} =T_{\phi\psi^{-1}}$.
By construction, a linear map $T$ belongs to $ N\left(  G\right)$ if, and only if, 
$\emph{Max}_G \left(  x\right)=\emph{Max}_G \left(  T(x)\right)$ for every $x\in\Fqn$ and it follows that $N(G)$ is a subgroup of $GL(n,G)_q$.
\end{IEEEproof}

\begin{lemma}
The group $Aut\left(  \widetilde{G}\right)  N\left(  G\right)  $ is the semidirect product
$Aut\left(  \widetilde{G}\right)  \ltimes N\left(  G\right)  $.
\end{lemma}

\begin{IEEEproof}
It is clear that $Aut\left(   \widetilde{G}\right)  \cap N\left(  G\right)  =\left\{
Id\right\}  $, so all is needed is to prove that $N\left(  G\right)  $ is
normal. This follows straightforward from the definition of the action of
$Aut\left(   \widetilde{G}\right)  $, noting that $Max (T_{\phi}\circ T\circ T_{\phi }^{-1}(x))=Max (T(x))  $.
\end{IEEEproof}

We claim that actually $GL\left(  n,G\right)  _{q}=Aut\left(   \widetilde{G}\right)
\ltimes N\left(  G\right)  $. In order to prove it, we first need to prove
some preliminary results.
\begin{lemma}
Given $T\in GL\left(  n,G\right)_{q}$ and $e_i\in\beta$, there are $v_{j(i)}\in \mathrm{supp}(T(e_i))$ and   $S_i \in N(G)$ such that: \emph{(i)}  $\left\langle
	\mathrm{supp}\left(  T\left(  e_{i}\right)  \right)  \right\rangle _{G}%
	=\left\langle v_{j(i)}\right\rangle_{G}$; \emph{(ii)} $\mathrm{supp}(S_iT(e_i))$ is a MSG for $\mathrm{supp}(T(e_i))$.
\end{lemma}
\begin{IEEEproof}
	\emph{(i)}   Since $T\in GL(n,G)_q$ it follows there is $v_{j(i)}\in \mathrm{supp}(T(e_{i}))$ such that $v_{i}\in \mathrm{supp}(T^{-1}(e_{j(i)}))$,  and  since $T^{-1}\in GL(n,G)_q$ it follows that $\left\langle
	\mathrm{supp}\left(  T\left(  e_{i}\right)  \right)  \right\rangle _{G}%
	=\left\langle v_{j(i)}\right\rangle _{G}$.	\emph{(ii)} Let $\widetilde{X}$ be a MSG of $\mathrm{supp}(T(e_i))$. Item 	\emph{(i)} ensures  that $\widetilde{X} = \{v_{j(i)}\}$. The linear isometry defined by $S_i(e_j) = e_j $ if $j\neq j(i)$ and $S(T(e_i)) = e_{j(i)}$, satisfies the desired conditions.
\end{IEEEproof}

The previous Lemma ensures the existence of a map $\phi_{T}:V\rightarrow
V$, where $\phi_{T}\left(  i\right)$ is defined to be a vertex such that
$\left\langle \mathrm{supp}\left(  T\left(  e_{i}\right)  \right)  \right\rangle
_{G}=\left\langle v_{\phi_{T}\left(  i\right)  }\right\rangle _{G}$, for each $T\in GL\left(  n,G\right) _{q}$. This map is not necessarily unique, since $\phi_{T}\left(i\right)$ may be exchanged by another vertex in the same clique. This amounts to the choice of a MSG made in the proof of item (i) of the previous Lemma to determine the map $S_i$. Considering the family $\mathcal{S} = \{S_i; 1\leq i \leq n\}$, we get a well defined map $S_{\phi_T}$:
\begin{proposition}\label{phi}
The map $\phi_{T}\in Aut(\widetilde{G})$ and $S_{\phi_{T}}=S_{n}\circ
\cdots\circ S_{2}\circ S_{1}\circ T$.
\end{proposition}

\begin{theorem}
$GL\left(  n,G\right)  _{q}=Aut\left(  \widetilde{G}\right)  \ltimes N\left(  G\right)  $.
\end{theorem}
\begin{IEEEproof}
It follows from Proposition \ref{phi} and the fact that $S_{\phi_{T} }\in Aut\left(   \widetilde{G}\right)  $, each $S_{i}\in N\left(  G\right)  $ and $T=S_{1}^{-1}\circ\cdots\circ S_{n}^{-1}\circ
S_{\phi_{T}}$.
\end{IEEEproof}

\section{G-Canonical Decomposition of linear codes for hierarchical graphs and its packing radius}\label{candec}

 As mentioned in Section \ref{sec:canonical}, the reduced canonical form $G'(V',E',L')$ of a graph $G(V,E)$ determines a partial order over $V'$, a poset, which we denote by $P_G=(V',\preceq_G)$. We may write $\preceq$ instead of $\preceq_G$, if no confusion may arise.

In the case the graph has no circuit, that is, in case $G$ defines a poset, the so-called hierarchical posets play an exceptional role,
since many of the known properties of codes, including MacWilliams Identity and Extension properties, hold for a poset metric  if and only if the poset is hierarchical (see \cite{KimOh,Machado,Oh}). Many of those results, originally proved for the usual Hamming metric, depends
essentially on the action of the group of linear isometries being transitive on
spheres centered at $0$.
We shall derive many similar properties for graphs for which the reduced canonical form determines a hierarchical poset. We will generalize the $P$-canonical decomposition for hierarchical posets, (see \cite{felix}), proving that each linear code  $\mathcal{C}\subset\Fqn$ is equivalent - up to linear isometry - to a code that may be expressed as a product of codes, each one having its support contained in a different level of the poset.  

%

We start giving proper definitions.  Let $G^{\prime}=G^{\prime}\left(  V^{\prime},E^{\prime},L^{\prime}\right)  $ be the
reduced canonical form of $G\left(  V,E\right)  $. Considering   $G'$ as a poset, it is naturally decomposed into levels. A \emph{chain} in $G'$ is a subset $I\subset V'$ such that given $u,v\in I$, either $u\preceq v$ or $v\preceq u$. The \emph{lenght} of a chain is just its cardinality. The \emph{height} $h_{G'}(v)$ of an element $v\in V'$ is the maximal length of a chain that has $v$ as a maximal element. The \emph{height } of $G'$ is $h_{G'}=\max\{h_{G´}(v); v\in V'\}$. The \emph{$i$-th level} $V'_i$ of  $G'$   is the set of elements of height $i$:
$V'_i=\{v\in V';h_{G'}(v)=i\}$. The level structure of $G'$ induces a similar structure in the original graph $G$: $h_G(v):=h_{G'}(\pi (v))$ and $V_i=\{v\in V;h_{G}(v)=i\}$.

 The set $V^{\prime}$ is decomposed as a disjoint union of its levels,  $V^{\prime}=V'_1 \cup\cdots\cup V'_{h_{G'}}$,
 called the \emph{level structure} of the graph $G^{\prime}$. A similar decomposition holds for the original graph $G$: $V=V_1 \cup\cdots\cup V_{h_{G}}$, where the union is disjoint and $h_G=h_{G'}$ The graph $G$ and the reduced $G'$ are \emph{hierarchical} if  $u\in V'_i$, $i>1$ and $v\in V'_{i-1}$, then
$u\rightarrow v$.

Two linear codes $\mathcal{C},\mathcal{C}^{\prime}\subset \F_q^n$ are said to be \emph{$G$-equivalent} if there is $T\in GL(n, G)_{q}$ such that $T(\mathcal{C})=\mathcal{C}^{\prime}$.


\begin{defn}Let $G(V,E)$ be a graph with $h_G$ levels. A linear code $\mathcal{C}\subset\Fqn$  \emph{admits a} $G$\emph{-canonical decomposition} if is $G$-equivalent to a linear code $\tilde{\mathcal{C}}=\mathcal{C}_1\oplus\cdots\oplus\mathcal{C}_{h_G}$, where   $\mathrm{supp}(\mathcal{C}_i)\subset V_i$.
\end{defn}

\begin{lemma}\label{lema_decomposicao} Let $G(V,E)$ be a  hierarchical graph with $h_G$ levels. Let $\mathcal{C}\subset\Fqn$ be a linear code  with $\mathrm{supp}(\mathcal{C})\subset V_i$, for some $i \leq h_G$. Consider $x \in \F_q^n$ such that $Max_G(x)\subset V_i$ and $\tilde{x}\not\in\mathcal{C}$. Then, $\mathcal{C}\oplus \mathrm{span}\{ x\}$ and $\mathcal{C}\oplus \mathrm{span}\{ \tilde{x}\}$ are $G$-equivalent.
\end{lemma}
\begin{IEEEproof}
Let $\{f_1, \ldots, f_k\}$ be a basis of $\mathcal{C}$. Since $\tilde{x}\not\in\mathcal{C}$, then $\alpha = \{\tilde{x}, f_1, \ldots, f_k\}$ is a basis of $\mathcal{C}\oplus  \mathrm{span}\{\tilde{x}\}$. We extend $\alpha$, by canonical vectors, to a basis $\beta=\{\tilde{x}, f_1, \ldots, f_k, e_{j_1}, \ldots, e_{j_r}\}$  of $\Fqn$. Since $Max_G(x)\subset\Gamma_G^i$ and $\tilde{x}\not\in\mathcal{C}$, we have that also $x\not\in \mathcal{C}$ and consequently $\alpha ' = \{x, f_1, \ldots, f_k\}$ is a basis of $\mathcal{C}\oplus \mathrm{span}\{x\}$ and   $\beta '=\{x, f_1, \ldots, f_k, e_{j_1}, \ldots, e_{j_r}\}$  of $\Fqn$. Let $T:\Fqn\rightarrow\Fqn$ be the linear map  defined by $T(x) = \tilde{x}$, $T(f_i)=f_i$ and $T(e_{j_k}) = e_{j_k}$. Clearly $T(\mathcal{C}\oplus \mathrm{span}\{ x\})=T(\mathcal{C}\oplus \mathrm{span}\{ \tilde{x}\})$.
By construction, $Max(y)=Max(T(y))  $ for every $y\in\Fqn$, hence  $T$ is an isometry.
\end{IEEEproof}
\begin{remark}
	The $G$-isometry $T$, constructed in the previous lemma, satisfies $\mathrm{supp}(T(\mathcal{C}\oplus span\{x\})\subset V_i$ and $T(y) = y$, for $\mathrm{supp}(y)\subset[n]\backslash V_i$. These properties will be used in the proof of Theorem \ref{thm:canonical_decomposition}.
\end{remark}

Given $x\in\Fqn$, the $i$-th \textit{projection} $\widehat{x}^i\in\Fqn$ is defined by $\widehat{x}_j^i= x_j$ if $j\in V_i$ and $\widehat{x}_j^i=0$ otherwise.
\begin{theorem}\label{thm:canonical_decomposition}
Let $G (V, E)$ be a graph. The poset $P_G$ is hierarchical if, and only if, any linear code $\mathcal{D}$ admits a $G$-canonical decomposition.
\end{theorem}
\begin{IEEEproof}
First, suppose that  $P_G$ is a hierarchical poset with $h_G$ levels. If  $dim(\mathcal{D})= 1$, it is enough to use Lemma \ref{lema_decomposicao}, considering  $\mathcal{C}=\{0\}$.

Suppose the result holds for linear codes with dimension smaller than $k$ and let $\mathcal{D} = span \{x_1\}  \oplus span\{x_2, \ldots, x_k\}$ be a $k$-dimensional code. The induction hypothesis ensures that, for $\mathcal{D}' = span\{x_2, \ldots, x_k\}$, there is a linear isometry $T'$ such that $\displaystyle T'(\mathcal{D}')=\oplus_{i=1}^l \mathcal{D}_i'$ and $\mathrm{supp}(\mathcal{D}_i')\subset V_i$. Since $T'(x_1)\not\in T'(\mathcal{D}')$, then $span\{ T'(x_1) \} \cap T'(\mathcal{D}')=\{0\}$ and there exists a level $i$ such that $\widehat{T'(x_1)}^i\not\in \mathcal{D}_i'$. Denote by $i_0$ the maximal level with this property and let $y$ be defined by $y_i = T'(x)_i$ if $i\in V_j$ with $j\leq i_0$ and $y_i = 0$ otherwise.  Thus we find that  $T'(\mathcal{D}) = span\{y\}  \oplus T'(\mathcal{D'})$. Considering $\mathcal{C} = \mathcal{C}_{i_0}'$, Lemma \ref{lema_decomposicao} ensures there is  $T\in GL(n,G)_{q}$ such that $T(\mathcal{D}_i') = \mathcal{D}_i'$ for $i\neq i_0$ and $\mathrm{supp}(T(span\{y\}  \oplus \mathcal{D'}_{i_0})) = \mathrm{supp}(span\{ \tilde{y} \}\oplus \mathcal{D'}_{i_0})\subset V_{i_0}$. Therefore, if $\mathcal{D}_i = \mathcal{D}_i'$ for $i\neq i_0$ and $\mathcal{D}_{i_0} = span\{\tilde{y} \}\oplus \mathcal{D'}_{i_0}$, then $T(T'(\mathcal{D})) = \oplus_{i=1}^l\mathcal{D}_i$ is a linear code $G$-equivalent to $\mathcal{C}$.

On the other hand, suppose that $P_G$ is not hierarchical and let $i\in[h_G]$ be the lowest level of $P_G$ for which there are $a \in V_i'$ and $b \in V_{i+1}'$ such that $a\not\preceq b$.
Consider $j\in \pi^{-1}(a)$ and $k\in \pi^{-1}(b)$. The linear code $\mathcal{C} = span\{ e_j+e_k\}$ cannot be $G$-equivalent to a canonically decomposed code  $\tilde{\mathcal{C}}$. Indeed, it follows from Proprosition \ref{phi} that any linear $G$-isometry $T\in GL(n, G)_{q}$ induces an automorphism $\Phi_T: V\mapsto V$. Thus, the closures $\langle \mathrm{supp}(T(e_j))\rangle_G$ and $\langle \mathrm{supp}(T(e_k))\rangle_G$ are generated by $\Phi_T(j) \in V$ and $\Phi_T(k)\in V$ respectively.  Moreover, since $\pi(j)\not\preceq \pi(k)$, we have that $\pi(\Phi_T(j))\not\preceq \pi(\Phi_T(k))$. It follows that $Max_G(T(span\{ e_j + e_k\})) \supset \{\Phi_T(j), \Phi_T(k)\}$ is not contained in a single level. Since $dim(\mathcal{C}) = 1$ and $T\in GL(n, G)_{q}$ is taken arbitrarily, we show that $\mathcal{C}$ does not admit a $G$-canonical decomposition.
\end{IEEEproof}

	\begin{remark}\label{gbasis}
		In the proof of the previous theorem, we constructed a map $T$, considering a basis $\beta=\{x_1, \ldots, x_k\}$ of $\mathcal{D}$. For the purpose of this Theorem, the choice of the basis is immaterial. For future purpose (Lemma \ref{lema_can_dec}), it is worth to note that the choice may be done in such a way that the linear isometry $T$ (which maps $\mathcal{D}$ into its $G$-canonical decomposition), restricted to $\beta$ is defined by $T(x_i) = \tilde{x_i}$. It follows that if $P_G$ is hierarchical, given a code $\mC$ it is possible to find a basis $\beta =\{x_1,x_2,\ldots ,x_k\} $ such that, when considering only the maximal components of each $x_i$, we get a basis $\tilde{\beta}=\{\tilde{x}_1,\tilde{x}_2,\ldots ,\tilde{x}_k\}$ such that $\SU (\tilde{x}_i)$ is contained in a single level of $P_G$ and $\tilde{\beta}$ generates a code $\mathcal{D}$ that is a $G$-canonical decomposition of  $\mC$.
	\end{remark}
Given a  code $\mathcal{C}\subset\Fqn$, we define its  \emph{minimal} $G$\emph{-distance} as $d(\mathcal{C}):=\min\{d_G(x,y);x,y\in \mathcal{C},x\neq y\}$. It is clear that for a linear code the minimal distance is the minimal weight $\min\{w_G(x);x\in \Fqn, x\neq 0\}$. The \emph{packing radius} $R(\mathcal{C})$ of the code is the maximal radius of disjoint balls centered at codewords:
\[
R(\mathcal{C}):=\max \{r\in [n]\cup \{0\}; B_r(x)\cap B_r(y)=\emptyset, \forall x,y\in \mathcal{C},x\neq y \} .
\]

In general, the packing radius is not determined by the minimal distance. Consider, for example, the graph with vertices set $V=\{v_1,v_2,v_3\}$ with the single edge $(v_3,v_1)$. Considering the one-dimensional codes $\mathcal{C}_1=\{000,110\}$ and $\mathcal{C}_2=\{000,001\}$ we get that $d(\mathcal{C}_1)=d(\mathcal{C}_2)=2$, while $R(\mathcal{C}_1)=0$ and $R(\mathcal{C}_2)=1$.

Let us consider the particular case where $G^{\prime}$ is hierarchical and $L^{\prime
}:V^{\prime}\rightarrow\mathbb{N}$ is constant on each level $V_{i}^{\prime}$,
let us say assuming the value $L^{\prime}\left(  i\right)$. In this situation, we have that $R(\mathcal{C})=R(d(\mathcal{C}))$, i.e., the packing radius $R\left(  \mathcal{C}\right)  $ of a linear code $\mathcal{C}$ is determined by its minimal distance $d\left(  \mathcal{C}\right)  $ as follows:

\begin{proposition}\label{weight}
Suppose that the reduced canonical form $G'(V',E',L')$ of $G$ is hierarchical and that $L'$ is constant and equal to $L'(i)$ on each level $V'_i$. Let $x\in \mathcal{C}$ be a codeword of minimal $G$-weight and let $k_0$ be the minimal level of $V^{\prime}$ intersected by $\mathcal{C}$: $k_0=\min \left\{ i; \mathrm{supp}(x)\cap V'_i\neq\emptyset ,x\in \mathcal{C}\right\}$. Then,
\[
R\left(  \mathcal{C}\right)  =\left\lfloor \dfrac{|\pi \left(\mathrm{supp} (x)\right)| - 1}{2 }\right\rfloor L^{\prime}\left(  k_0\right)
+ \sum_{i=1}^{k_{0}-1}  |V'_i|  L^{\prime}\left(
i\right)  \text{,}%
\]
where $X$ is a MSG for $\langle \mathrm{supp}(x)\rangle_G$.
\bigskip
\end{proposition}

\begin{IEEEproof}
Let us assume that $\mathcal{C}$ is  $G$-canonical decomposed as  $\mathcal{C}_{k_0}\oplus \mathcal{C}_{k_0+1} \oplus\cdots\oplus\mathcal{C}_{h_G}$. Given $x\in \mathcal{C}_{k_0}$, since $L'$ is assumed to be constant on each level, it is immediate to check that
\[
w_G(x) = |(\pi\left(\mathrm{supp}(x)\right)|L'(k_0)) +  \sum_{i=1}^{k_{0}-1}|V'_i|L'(i).
\]
Considering the $G$-canonical the $G$-canonical decomposition, we have that any codeword  $x\in\mathcal{C}_{k_0}$ is a cleared codeword hence the minimal distance is attained by such a vector $x_0$ with $|\pi (\mathrm{supp}(x_0))|$ minimal. Let us denote $r_0=|\pi (\mathrm{supp}(x_0))|$. Given two codewords $x,y\in\mathcal{C}_{k_0}$, the minimality of $r_0$ ensures that $| \pi (\mathrm{supp}(x))\cap \pi (\mathrm{supp}(y))| \geq \left\lfloor r_0 \right\rfloor $  and hence $d_G(x,y)\geq \left\lfloor r_0 \right\rfloor  L'(k_0) +  \sum_{i=1}^{k_{0}-1}|V'_i|L'(i)$. To show that this radius can not be increased, we choose a subset $I\subset \pi (\mathrm{supp}(x_0))$ with $|J|= \left\lfloor r_0 \right\rfloor +1$  $w_G(x)\leq $ and define $y=(y_1,\ldots,y_n)\in\mathcal{C}_{k_0}$ by
$y_j=x_j$ if $j\in J$ and $y_j=0$ otherwise. It follows that
\[
d_G(x_0,y)   =\left( \delta -1 \right) L^{\prime}\left(  k_0\right)
+ \sum_{i=1}^{k_{0}-1} |V'_i| L^{\prime}\left(
i\right)
\text{\hspace{6pt} and \hspace{6pt} }
w_G(y)  =\left( \delta +1 \right) L^{\prime}\left(  k_0\right)
+ \sum_{i=1}^{k_{0}-1}|V'_i|  L^{\prime}\left(
i\right).
\]
where  $\delta=\left\lfloor \dfrac{|\pi \left( \mathrm{supp} (x)\right)| - 1}{2 }\right\rfloor$. Denoting $\rho = \left( \delta +1 \right) L^{\prime}\left(  k_0\right)
+ \sum_{i=1}^{k_{0}-1} |V'_i|  L^{\prime}\left(
i\right)$  we have that $d_G(x_0,y)<\rho$ and $w_G(y)=\rho$, that is,   $y\in B_{\rho}(0)\cap B_{\rho }(x_0)$, hence $R(\mathcal{C})=  \delta  L^{\prime}\left(  k_0\right)
+ \sum_{i=1}^{k_{0}-1}|V'_i|  L^{\prime}\left(
i\right)$.
\end{IEEEproof}

From Proposition  \ref{weight}, $R(\mathcal{C})=R(d_G(\mathcal{C}))$.  We remark that this
situation, whether the packing radius is determined by the minimal distance, is
unusual but very helpful, since even for codes of dimension
$1$, determining the packing radius is, in general, an NP-hard problem (see
\cite[Section 4]{Oliveira}).

\section{The MacWilliams Identity and Extension Theorem}\label{mac}

In this section, we introduce the MacWilliams Identity and the MacWilliams Extension Property in the context of graph-metrics. The condition established in Theorem~\ref{thm:canonical_decomposition} for a graph $G(V, E)$ to admit a $G$-decomposition is not sufficient to ensure the MacWilliams properties. Indeed, consider the metric induced by the graph $G ([4], E=\{(3,4), (4,3)\})$. 
Let  $\mathcal{C}_1=\{0000, 1100\}$ and $\mathcal{C}_2=\{0000, 0011\}$ be two one-dimensional binary codes. It is clear that $\mathcal{C}_1$ is isometric to $\mathcal{C}_2$, since $w_G(1100)=w_G(0011)=2$. The dual codes $\mathcal{C}_1^{\perp}$ and $\mathcal{C}_2^{\perp}$ are generated by the sets $\{ 1100,1110,1101\}$ and $\{ 0011,1011,0111\}$ respectively. Direct calculations shows that  $W^G_{\mathcal{C}_1^{\perp}}(X) =1+4X^2+3X^4$ and $W^G_{\mathcal{C}_2^{\perp}}(X)) =1+2X+2X^2+2X^3+X^4$.
Therefore, the MacWilliams Identity does not hold in full generality. It is also easy to check (considering the description of $GL(n,G)_q$ given in Section \ref{sec:coding})  that no linear isometry  can map $1100$ into $0011$, i.e., MacWilliams Extension Property also does not hold in the general case.  Our aim now is to find sufficient conditions to ensure the existence of a MacWilliams identity. Surprisingly, as we shall see, the Extension Property is strictly stronger than the MacWilliams Identity.

We start with a definition that will be crucial for both the cases.
\begin{defn}
	Let $G'(V', E')$ be the reduced canonical form of $G$ and let $V_1'\cup \cdots \cup V'_{h_G}$ be the level decomposition of $V'$. We say that $G$ satisfies the \textit{Unique Decomposition Property} (UDP)  if $S,S^{\prime}\subset V_i'$ are sets satisfying $ \sum_{a\in S} L'(a) = \sum_{b\in S^{\prime}} L'(b),$ then there is a bijection $g : S \rightarrow S^{\prime}$ such that $L'(a) = L'(g(a))$ for all $a\in S$.
\end{defn}
\begin{proposition}\label{hier}
Suppose that the reduced canonical form $G^{\prime}=\left(V^{\prime},E^{\prime
	},L^{\prime}\right)$ of $G$ is a hierarchical poset. The group of linear isometries $GL\left(  n,G\right)_{q}$ acts transitively on the
	spheres of $\left(  \mathbb{F}_{q}^{n},d_G \right)  $ if, and only if, $G$ satisfies the UDP.
\end{proposition}

\begin{IEEEproof}
	Given $x\in\mathbb{F}_q^n$, let $\widetilde{X}$ be a MSG for $\langle\mathrm{supp}(x)\rangle_G$. Since $G^{\prime}$ is hierarchical, all elements of $\widetilde{X}$ belong to the same level, let us say $\widetilde{X}\subset V'_{k_0}$. Then we find that
	\[
	w_G(x) =  \sum_{v\in \widetilde{X}} L^{\prime }(v)  + \sum_{i=1}^{k_0-1} \sum_{v\in V'_{i}}  L^{\prime}(v)
	\]
	and the result  follows from Proposition \ref{phi}.
\end{IEEEproof}
\subsection{The MacWilliams Identity}

We denote  $A_i^G(\mathcal{C})=|\{ c\in \mathcal{C}:w_G(c)=i\}|$ and define the $G$-\emph{weight enumerator} of a code $\mathcal{C}$ as the polynomial
$$W^G_{\mathcal{C}}(X)=\sum_{i=0}^n A^G_i(\mathcal{C})X^i.$$

Given a graph $G(V,E)$, let us denote by  $\overline{G}(V,{\overline{E}})$ its \textit{reverse graph}.
It is easy to see that a closure $J\in\mathcal{I}(G)$ if, and only if, its complement $J^c\in \mathcal{I}(\overline{G})$. Furthermore, $A\subset V$ is a circuit in $G$ if, and only, if, $A$ is also a circuit in $\overline{G}$. Therefore, the reduced canonical form $\overline{P}_G(V',\preceq_{\overline{P}_G})$ of $\overline{G}$ is the dual of $P_G(V', \preceq_{P_G})$.



\begin{defn}(The MacWilliams Identity) A graph $G(V,E)$ admits a MacWilliams Identity if for every linear code $\mathcal{C}\subset\Fqn$, the $G$-weight enumerator $W_{\mathcal{C}}^G(X)$ of $\mathcal{C}$ determines the $\overline{G}$-weight enumerator $W_{\mathcal{C}^{\perp}}^{\overline{G}}(X)$ of the dual code $\mathcal{C}^{\perp}$.
\end{defn}

In order to characterize the graphs that admit the MacWilliams Identity, we briefly introduce some facts about additive characters that will be used in  a manner similar  to MacWilliams' original approach, later adapted by Choi et al \cite{macwilliams-equivalence} to the poset case. An \textit{additive character} $\chi$ of $\mathbb{F}_q$ is a nontrivial homomorphism of the additive group $\mathbb{F}_q$ into the multiplicative group of complex numbers with 1-norm. The next  lemma is well known.

\begin{lemma}\label{caract} Let $\mathcal{C}\subset\mathbb{F}_q^n$ be a linear code and let $\chi$ of $\mathbb{F}_q$ be an additive character. Then,
$\sum_{x\in \mathcal{C}}\chi(x\cdot y) = \mathcal{C}$ if  $ y\in\mathcal{C}^{\perp}$ and $0$ otherwise.
\end{lemma}

\begin{lemma}\label{coeficiente}Let $G(V,E)$ be a finite graph. Given a linear code $\mathcal{C}$ of $\mathbb{F}_q^n$, then
$$A_i^G(\mathcal{C})= \frac{1}{|\mathcal{C}^{\perp}|}\sum_{1\leq j\leq n}\sum_{x\in\mathcal{C}^{\perp}\cap \overline{S}_j}\sum_{y\in S_i}\chi(x\cdot y),$$
where $S_i$ and $\overline{S}_j$ are the spheres of radii $i$ and $j$ considering the metric induced by the graphs $G$ and $\overline{G}$, respectively, i.e.,
$S_i = \{x\in  \mathbb{F}_q^n : w_{G}(x) = i\}$ and  $\overline{S}_j = \{x\in  \mathbb{F}_q^n : w_{\overline{G}}(x) = j\}$.
\end{lemma}
\begin{IEEEproof}
First, note that $A_i^G(\mathcal{C}) = \left|\mathcal{C}\cap S_i\right| = \sum_{x\in \mathcal{C}\cap S_i}1$. Lemma \ref{caract} implies that
$$A_i^G(\mathcal{C}) =  \sum_{y\in S_i}\frac{1}{|\mathcal{C}^{\perp}|}\sum_{x\in \mathcal{C}^{\perp}}\chi(x\cdot y) =\frac{1}{|\mathcal{C}^{\perp}|} \sum_{x\in \mathcal{C}^{\perp}}\sum_{y\in S_i}\chi(x\cdot y) =\frac{1}{|\mathcal{C}^{\perp}|}\sum_{1\leq j\leq n}\sum_{x\in\mathcal{C}^{\perp}\cap \overline{S}_j}\sum_{y\in S_i}\chi(x\cdot y). \vspace{-22pt} $$
\end{IEEEproof}

\begin{lemma}\label{caracteres}A directed graph $G(V,E)$ admits the MacWilliams Identity if, given $i, j\in [n]$, and  $x, x' \in S_i$, then $$\sum_{y\in\overline{S}_j} \chi(x\cdot y) = \sum_{y\in\overline{S}_j} \chi(x'\cdot y).$$
\end{lemma}
\begin{IEEEproof}
Let $\mathcal{C}_1, \mathcal{C}_2 \subset \mathbb{F}_q^n$ be linear codes such that $W_{\mathcal{C}_1}^G(X) = W_{\mathcal{C}_2}^G(X)$. Since $G$ satisfies the first condition we have that  $p_{ij} := \sum_{y\in\overline{S}_j}\chi(x\cdot y)$ does not depend on the choice of $x\in S_i$. Therefore, $W^{\overline{G}}_{\mathcal{C}_1^{\perp}}(X)\!=\! W^{\overline{G}}_{\mathcal{C}_2^{\perp}}(X)$, since, from Lemma \ref{coeficiente},
$$A_j^{\overline{G}}(\mathcal{C}_k^{\perp})=\frac{1}{|\mathcal{C}_k|}\sum_{1\leq i\leq n}\sum_{x\in\mathcal{C}_k\cap S_i}p_{ij} = \frac{1}{|\mathcal{C}_k|}\sum_{1\leq i\leq n}A_i^G(\mathcal{C}_k)p_{ij}, \text{ for } k\in \{1,2\}. \vspace{-22pt}$$
\end{IEEEproof}


\begin{lemma}\label{p_ij}
		Let $G (V,E)$ be a graph with $h_G=1$ and let $P_G=(V',\preceq)$ be its reduced canonical form (considered as a poset). Given $I, J \in \mathcal{I}(G)$, let  $S_{I} =  \{x\in \mathbb{F}_q^n : \langle \mathrm{supp}(x)\rangle_G = I\}$. Then, for   $x \in S_{I}$ we have that,
\[
\sum_{y\in S_{J^{c}}}\chi(x\cdot y) =(-1)^{|\pi( I \cap J^c)|}\hspace{-.2cm} \prod_{i\in\pi (I^c \cap J^c)}(q^{L(i)}-1)\prod_{j\in\pi(J^c)}(q^{L(j)}-1)^{|\pi(J^c)|-1}.
\]
\end{lemma}

\begin{IEEEproof}
Given $x\in\Fqn$, the vector $\overline{x}^i\in\Fqn$ is defined by $\overline{x}_j^i= x_j$ if $j\in \pi^{-1}(i)$ and $\overline{x}_j^i=0$ otherwise.	In order to simplify the notation, we assume that $V'= \{1, \ldots, m\}$, for an integer $m\leq |V|$. Thus,
	\begin{align*}
	\sum_{y\in S_{J^{c}}}\chi(x\cdot y) = \sum_{y\in S_{J^{c}}}\chi\left(\sum_{i=1}^m \overline{x}^i\cdot \overline{y}^i\right)
	= \sum_{y\in S_{J^{c}}}\prod_{i\in \pi(J^c)}\chi\left(\overline{x}^i\cdot \overline{y}^i\right) \prod_{i\in \pi(J)}\chi\left(\overline{x}^i\cdot \overline{y}^i\right)
	= \prod_{i\in \pi(J^c)}\sum_{y\in S_{J^{c}}}\chi\left(\overline{x}^i\cdot \overline{y}^i\right),
	\end{align*}
where in the first equality we consider the fact that $\chi$ is a group homomorphism and we separate the product in the $J$ and $J^c$ parts and the second follows from the fact that $\overline{x}^i\cdot \overline{y}^i=0$ for $i\in\pi (J)$ and $\chi(0)=1$.	We note that, given $\overline{y}^1+ \cdots+ \overline{y}^m \in S_{J^c}$ we have  that   $\overline{y}^i \in \mathbb{F}_q^{L(i)}\backslash\{0\}\subset \mathbb{F}_q^n$ if  $i\in \pi(J^c)$ and $\overline{y}^i= 0$, otherwise. It follows that
	\[
	\sum_{y\in S_{J^{c}}}\chi\left(\overline{x}^i\cdot \overline{y}^i\right) = \sum_{\overline{y}^1+ \cdots+ \overline{y}^m\in S_{J^{c}}}\chi\left(\overline{x}^i\cdot \overline{y}^i\right)
	=\prod_{j\in\pi(J^c)\backslash\{i\}}(q^{L(j)}-1)\sum_{\overline{y}^i\in \mathbb{F}_q^{L(i)}\backslash\{0\}\subset \mathbb{F_q}^n}\chi\left(\overline{x}^i\cdot \overline{y}^i\right).
	\]
	Furthermore, from Lemma \ref{caract}, we have that $ \sum_{y\in \mathbb{F}_q^{L(i)}\backslash\{0\}}\chi\left(y\cdot z\right) =  	q^{L(i)}-1$ if $z=0$ and equals $-1$ otherwise.
	Hence,
	\begin{align*}
	\sum_{y\in S_{J^{c}}}\chi(x\cdot y)
	&= (-1)^{|\{i\in V': \mathrm{supp}(\overline{x}^i)\cap J^c\neq \emptyset\}|} \hspace{-1cm}\prod_{i\in\{j\in V' : \overline{x}_j=0\hspace{0.05cm}, \pi^{-1}(j)\subset J^c\}}\hspace{-1cm} (q^{L(i)}-1)\prod_{j\in\pi(J^c)}(q^{L(j)}-1)^{|\pi(J^c)|-1}\\
	&	= (-1)^{|\pi( I \cap J^c)|}\hspace{-.2cm} \prod_{i\in\pi (I^c \cap J^c)}(q^{L(i)}-1)\prod_{j\in\pi(J^c)}(q^{L(j)}-1)^{|\pi(J^c)|-1}.
	\end{align*}	
where the second equality follows from the fact that $ |\{i\in V': \mathrm{supp}(\overline{x}^i)\cap J^c\neq \emptyset\}| = |\{i\in V'\!:\! \pi^{-1}(i)\subset I \cap J^c\}|\! =\! |\pi( I \cap J^c)|$.
\end{IEEEproof}
 In the next theorem, we will use a variation of the previous lemma, considering not only the closed set $J^c$, but the family $\overline{J^c}$ of all closed sets with the same cardinality of  $J^c$: Given $I, J \in \mathcal{I}(G)$ and $x\in S_I$, then
\begin{align*}
&\sum_{y\in S_{\overline{J^{c}}}}\chi(x\cdot y)= \sum_{K^c \in \overline{J^{c}}}\sum_{y\in S_{K^{c}}} \chi(x\cdot y)
= \sum_{K^c \in \overline{J^{c}}} (-1)^{|\pi( I \cap K^c)|}  \prod_{i\in\pi(I^c \cap K^c)}(q^{L(i)}-1)\prod_{j\in\pi(K^c)}(q^{L(j)}-1)^{|\pi(K^c)|-1}.
\end{align*}


\begin{theorem} \label{thm:mac_1-level}Let $G(V, E)$ be a graph and suppose that its reduced canonical form $P_G=(V',\preceq)$ is an anti-chain. The graph $G$ admits the MacWilliams Identity if, and only if, $G$ satisfies the UDP.
\end{theorem}
\begin{IEEEproof}	
	It is clear to see that $|I|= |J| \iff |I^c| = |J^c|$. Given $x,x^{\prime}\in S_i$, consider the closures $I_1 =\langle \mathrm{supp}(x) \rangle_G$ and $I_2=\langle \mathrm{supp}(x^{\prime})\rangle_G$. Thus, $\sum_{i\in\pi(I_1)}L(i) = \sum_{i\in\pi(I_2)}L(i)$. If $G$ satisfies the UDP, then there is a bijection $g: \pi(I_1) \rightarrow \pi(I_2)$ such that $L(i)= L(g(i))$. Furthermore, since $\sum_{i\in\pi(I_1^c)}L(i) = \sum_{i\in\pi(I_2^c)}L(i)$, then $g$ may be extended to $V^{\prime}$. It implies that there is a $G$-automorphism $\varphi\in Aut(G)$ such that $\varphi(I_1) = I_2$.
	
	Given $J\in\mathcal{I}(G)$, consider $j=|J|$. From Lemma \ref{p_ij} it follows that
	$	\sum_{y\in S_j}\chi(x\cdot y)= \sum_{y\in S_j}\chi(x^{\prime}\cdot y).$
	%
Since $h_G=1$, the conditions 1) and 2) of Lemma \ref{caracteres} are equivalent and, then $G$ admits the MacWilliams Identity.

On the other hand, let $S, S' \subset V'$ be minimal sets such that: \emph{(i)}  $\sum_{i\in S}L'(i) = \sum_{i\in S'}L'(i)$; \emph{(ii)} there is no bijection $g: S \rightarrow S'$  such that $L'(i)=L'(g(i))$.
Consider the linear codes  $\mathcal{C}_1 = span \{x\}$ and $\mathcal{C}_2 = span \{y\}$, where $x = \sum_{i\in S}e_i$ and $y = \sum_{i\in S'}e_i$. By construction, the $G$-weight enumerators of $\mathcal{C}_1$ and $\mathcal{C}_2$ are equal. Furthermore, by the minimality of $S$ and $S'$, $\mathrm{supp}(\mathcal{C}_1)\cap \mathrm{supp}(\mathcal{C}_2)=\emptyset$ and $L'(i)\neq L'(j)$, for any $i\in S$ and $j\in S'$. It follows that there exists $i_0\in S\cup S'$ such that $L'(i_0)\leq L'(j)$, for each $j\in S\cup S'$. Suppose $i_0 \in S$, then there is $z\in\Fqn \setminus \mathcal{C}_1^{\perp}$ such that $\mathrm{supp}(z)\subset \pi^{-1}(i_0)$,  what implies $A_{w_G(e_i)}(\mathcal{C}_1^{\perp}) < A_{w_G(e_i)}(\mathcal{C}_2^{\perp})$.
\end{IEEEproof}
\begin{lemma}\label{lema_can_dec}
Let $G(V,E)$ be a graph and suppose that $P_G$ is a hierarchical poset with $l$ levels. Let $\mathcal{C}$ be a linear code and $\mathcal{C}_1\oplus \cdots \oplus\mathcal{C}_l$ its $G$-canonical decomposition. Let $\mathcal{D}_i=\{y\in \mathcal{C}_i^{\perp}: \mathrm{supp}(y)\subset V_i \}$ and $\mathcal{D}=\mathcal{D}_1\oplus \cdots \oplus \mathcal{D}_l$. Then, $\mathcal{C}^{\perp}$ and $\mathcal{D}$ are $\overline{G}$-equivalent.
\end{lemma}
%
%
%

\begin{IEEEproof}
Let $\alpha = \{x_1,x_2,\ldots ,x_r\}$ be a basis of $\mathcal{C}$ such that $\{\tilde{x}_1,\tilde{x}_2,\ldots ,\tilde{x}_r\}$ generates a $G$-canonical decomposition $\mathcal{C}_1\oplus \cdots \oplus\mathcal{C}_l$, as ensured by Remark \ref{gbasis}. Since $P_G$ is assumed to be hierarchical, so is $P_{\overline{G}}$ and it follows, again by Remark \ref{gbasis}, that $\mC^{\perp}$ admits a basis  $\beta = \{y_1,y_2,\ldots ,y_{n-r}\}$ such that  $\tilde{\beta}=\{\tilde{y}_1,\tilde{y}_2,\ldots ,\tilde{y}_{n-r}\}$ generates a code $\overline{G}$-equivalent to $\mathcal{C}^{\perp}$, that is a $\overline{G}$-canonical decomposition of $\mC^{\perp}$. We claim that this code is actually $\mathcal{D}$. Indeed, to conclude that, we need to prove $\tilde{x}_j \cdot \tilde{y}_k = 0$, for all $1\leq j\leq r$ and $1\leq k \leq n-r$. Actually, since all those vectors have the support in a single level, it is enough to prove $\tilde{x}_j \cdot \tilde{y}_k = 0$ for $\tilde{x}_j \in \tilde{\alpha}$ and $\tilde{y}_k\in\tilde{\beta}$ such that $\SU(\tilde{x}_j),\SU(\tilde{y}_k)\subset V_i$. Note that, if $\SU(\tilde{x}_j)\subset V_i$, then $\SU(x_j)\subset V_1 \cup V_{2}\cup \cdots \cup V_{i}$. Analogously, if $\SU(\tilde{y}_k)\subset V_i$, then $\SU(y_k)\subset V_i \cup V_{i+1}\cup \cdots \cup V_{l}$, since the metric is induced by the reverse graph $\overline{G}$. This implies that $x_j \cdot y_k= \tilde{x}_j \cdot \tilde{y}_k$ and since $x_j \cdot y_k=0$ we get that also $\tilde{x}_j \cdot \tilde{y}_k=0$.
	\end{IEEEproof}
\begin{theorem}
Let $G(V,E)$ be a graph and suppose that $P_G$  is a hierarchical poset. The graph $G$ admits the MacWilliams Identity if, and only if, $G$ satisfies the UDP.
\end{theorem}
\begin{IEEEproof}
	Suppose that the graph $G(V,E)$ has $h_G$ levels with $n_i$ elements in the $i$-th level. From Lemma \ref{lema_can_dec}, we may assume, without loss of generality, that a linear code $\mathcal{C}\subset\F_q^n$ is already in a $G$-canonical form $\mathcal{C}_1\oplus \cdots \oplus\mathcal{C}_{h_G}$ and its dual is equivalent to $\mathcal{C}^{\perp} = \mathcal{D}_1 \oplus \mathcal{D}_2 \oplus\cdots \oplus \mathcal{D}_l$, where $\mathcal{D}_i$ is defined as in Lemma \ref{lema_can_dec}.
	
	
	Furthermore, since the graph $G$ satisfies the Unique Decomposition Property, the graph induced on $V_i'$ also satisfies the Unique Decomposition Property and so, Theorem \ref{thm:mac_1-level} ensures that $W_{\mathcal{C}_i}(X)$ determines $W_{\mathcal{D}_i}(X)$. Since $W_{\mathcal{C}}^G(X)$ can be expressed in terms of the $W_{\mathcal{C}_i}(X)$'s (and similarly for $W_{\mathcal{C}^{\perp}}(X)$), we conclude the proof.
 $W^{\overline{G}}_{\mathcal{C}^{\perp}}(X)$ is completely determined by $W^{G}_{\mathcal{C}}(X)$.
\end{IEEEproof}

\begin{remark}
 As we saw, a poset-block metric can be obtained as a variation of a graph based metric by imposing $L_1(v)=1$ for every vertex $v$ in the canonical reduced form. In this situation,  the UDP condition is equivalent to demand that every block has the same size. For this fact, the condition we establish for the existence of a MacWilliams identity may be seen as a generalization of the conditions established in \cite{Jerry} for the case of poset-block metrics. We also remark that the UDP condition is  necessary and sufficient only by assuming that the graph is hierarchical. That is not the case for poset-block metrics.
\end{remark}

\subsection{The MacWilliams Extension Theorem}
\begin{defn}(\emph{The MacWilliams Extension Property})
	A metric space $(\mathbb{F}_q^n, d_G)$ \emph{ satisfies the MacWilliams Extension Property} if for any pair of linear codes $\mathcal{C}$ and $\mathcal{C}'$ and any linear map $t: \mathcal{C}\rightarrow\mathcal{C}'$ preserving the $G$-weight, there is a $G$-isometry $T\in GL(n, G)_{q}$ such that $T|_{\mathcal{C}}=t$.
\end{defn}

The same conditions on the graph $G$ that ensured the MacWilliams Identity (the canonical reduced form being an hierarchical poset and the UDP) are not sufficient to characterize those metric spaces $(\mathbb{F}_q^n, d_G)$  that satisfy the MacWilliams Extension Property. In fact, consider the graph $G([6]$, with set of edges $E= \{(1,2),(2,1), (3,4), (4,3), (5,6),(6,5)\})$. We consider the space $(\mathbb{F}_2^n, d_G)$ and the linear codes $
\mathcal{C}_1 = \{000000, 100010, 101000, 001010\}$ and
$ \mathcal{C}_2 = \{000000, 010010, 110011, 100001\}$.
The linear map $t: \mathcal{C}_1 \rightarrow\mathcal{C}_2$ defined by $t(100010) = 010010$ and $t(101000) =110011$  preserves the $G$-weight but any possible linear extension does not preserve the $G$-weight of the vector $100000$; hence $t$ cannot be extended to  $T\in GL(n, G)_{2}$.

This situation can be avoided by adding an extra condition.

\begin{defn}(\emph{Condition} $\Omega$) Let $G(V,E)$ be a directed graph and $G'(V',E',L)$ be its reduced canonical form. We say that   $G$ and $G'$ satisfy the condition $\Omega$ if, given an integer $k>1$, there are at most two elements $i, j\in V^{\prime}$ such that $L(i)= L(j) = k$.
\end{defn}

From here on, we assume that the field $\mathbb{F}_q$ is binary, that is, $q=2$.

Let us assume that the reduced canonical form  $G'$ of $G$ is a poset with a single level, that is, $E'=\emptyset$. Given a linear code $\mC\subset \mathbb{F}_2^n$. We define
$
I_j(\mC)=\{k\in \pi (\SU (\mC));L(k)=j \}.
$
When considering two codes $\mC_1$ and $\mC_2$, we shall write $I_j(\mC_i)=I_j^i$.

\begin{lemma}\label{ext_mac}
	Let $G$ be a directed graph satisfying the UDP and condition $\Omega$. Let $\mC_1$ and $\mC_2$ be linear codes in $(\mathbb{F}_2^n, d_G)$ and $t:\mC_1\rightarrow \mC_2$  a linear map preserving the $G$-weight. Then, $|I^1_j|=|I^2_j|$, for all $j\leq r=\max \{L(k); k\in V'\}$.
\end{lemma}
\begin{IEEEproof}
		Let us decompose the support of the codes as
	$
		\SU (\mC_i)=I^i_1\cup I^i_2 \cup \cdots \cup I^i_{r}.
	$
		We remark that some of the $I^i_j$ may be empty; however, once  $I^1_j=\emptyset$, we must have that also $I^2_j=\emptyset$. Indeed, suppose that $I^1_j\neq \emptyset$. This means there is $x\in\mC_1$ such that $\pi (\SU (x))\cap I^1_j\neq \emptyset$. The UDP ensures that  $\pi (\SU (t(x)))\cap I^2_j$ is also nonempty.
		
		Let us consider $2\leq j \leq r$. Suppose that  $|I^1_j|\leq |I^2_j|$, since $|I^i_j|\leq 2$ by hypothesis, then the possible values for $(|I^1_j|,|I^2_j) $ are $(0,0),(0,1),(0,2),(1,1),(1,2)$ and $(2,2)$. The cases  $(0,1)$ and $(0,2)$ cannot occur, since, as we just saw, $I^1_j=\emptyset$ \emph{iff} $I^2_j=\emptyset$. We need to discard the case $(|I^1_j|,|I^2_j)=(1,2) $. Let $x\in\mC_1$ be a vector such that $\pi(\SU (x))=I^1_j$. The UDP ensures there is no $y\in\mC_2$ such that $\pi(\SU (y))=I^2_j$. However, there must be $y,z\in\mC_2$ such that $\pi(\SU (y))\cup \pi(\SU (z))=I^2_j$ and then we have that either $\pi(\SU (y))=I^2_j$, or $\pi(\SU (z))=I^2_j$, or  $\pi(\SU (y+z))=I^2_j$, a contradiction.
		
		Now, we need to prove that $I^1_1=I^2_1$. We consider a set $X=\{x_1,x_2,\ldots ,x_s\}$ such that $\bigcup_{i=1}^s (\pi(\SU(x_i))\cap I_1^1)=I_1^1 ,$ with $s$ minimal. A careful use of the inclusion-exclusion principle and the UDP ensures that
		\[
		\left| \bigcup_{i=1}^s (\pi(\SU(x_i))\cap I_1^1) \right| =\left| \bigcup_{i=1}^s (\pi(\SU(t(x_i)))\cap I_1^2) \right| \leq |I^2_1|
		\]
		
		and 
	    	we get that $|I^1_1|\leq |I^2_1|$. A similar reasoning for the inverse $t^{-1}$ ensures that $|I^1_1|= |I^2_1|$.		
	\end{IEEEproof}


\begin{proposition}\label{1level}
	Let $G(V,E)$ be a directed graph and suppose that its canonical reduced form is a poset with a single level ( $h_G=1$). The metric space $(\mathbb{F}_2^n,d_G)$ satisfies the MacWilliams Extension Property if, and only if, $G$ satisfies the UDP and the Condition $\Omega$.
\end{proposition}
\begin{IEEEproof} First of all, we shall prove that the two stated conditions are sufficient.
	Let $\mathcal{C}_1$ and $\mC_2$ be two linear codes and let $t:\mC_1\rightarrow\mC_2$ be a linear map that preserves the $G$-weight. Given $x,y\in\mC_1$, the UDP ensures the existence of the bijections $g_1: \pi(\SU(x))\rightarrow \pi(\SU(t(x)))$ and  $g_2: \pi(\SU(y))\rightarrow \pi(\SU(t(y)))$. We claim that  it is possible to choose $g_1$ and $g_2$ in such a way that, if $i\in \pi(\SU(x)) \cap \pi(\SU(y))$, then $g_1(i) = g_2(i)$. Indeed, suppose that $g_1(i) \neq g_2(i)$ and consider the linear codes $\mC = span\{x,y\}$ and $\mC^{\prime} = span \{t(x), t(y)\}$. The only obstructions to $g_1$ and $g_2$ to satisfy this condition would be if either $|I_{L(i)}(\mC)|< |I_{L(i)}(\mC')|$ or $|I_{L(i)}(span\{x\})|< |I_{L(i)}(span\{t(x)\})|$, contradicting Lemma \ref{ext_mac}.
	Hence, there is  a bijection $\phi_t: \pi(\SU(\mC_1))\rightarrow\pi(\SU(\mC_2))$ such that $\phi_t: \pi(\SU(x))\rightarrow\pi(\SU(t(x)))$ is a bijection preserving the $L$-weight. Since $G$ satisfies the UDP and $$\sum_{i\in V^{\prime}\backslash\pi(\SU(\mC_1))} L(i)= \sum_{i\in V^{\prime}\backslash\pi(\SU(\mC_2))}L(i),$$ then $\phi_t$ can be extended to $\varphi:V^{\prime}\rightarrow V^{\prime}$.
	Given $x\in\mathbb{F}_2^n$, we write $x = x_1 + \cdots +x_m$, where $\pi(\SU(x_i)) \subset \{v_i\} \in V'$. It follows that $\mC_j \subset \mC_{j1} \oplus \cdots \oplus \mC_{jm}$, where $\mC_{ji}=\{x_i; x\in\mC_j\}$, for $j=1,2$. Let $\beta_{j} = \{x_{j1}, \ldots, x_{jk_j}\}$ be a basis of $\mC_{1j}$ and $\alpha_{\varphi(j)} = \{t(x)_{\varphi(j)1}, \ldots, t(x)_{\varphi(j)k_j}\}$ be a basis of $\mC_{2\varphi(j)}$. We consider $W_j$ to be the subspace of $\mathbb{F}_2^{n}$  isomorphic to $\mathbb{F}_2^{L(i)}$ such that $\pi(\SU(e_{jk})) = v_j$ and  we extend the basis $\beta_{j}$ of $\mC_{1j}$ to a basis $\beta_{j}^{\prime} = \{x_{j1}, \ldots, x_{jk_j},e_{j1}, \ldots, e_{jr_j}\}$ of  $W_j$.  It follows that, $\mathbb{F}_2^n = \bigoplus_{i=1}^m W_j$. Analogously, we can also extend $\alpha_{\varphi(j)}$ to $\alpha_{\varphi(j)}^{\prime}=\{t(x)_{\varphi(j)1}, \ldots, t(x)_{\varphi(j)k_j}, f_{\varphi(j)1}, \ldots, f_{\varphi(j)r_j}\}$ to another basis of $W_{\varphi(j)}$ . Let $T: \F_2^n \rightarrow \F_2^n	$
be the linear map defined by $T(x_{ij})= t(x)_{\varphi(i)j}$ and $T(e_{ij})= f_{\varphi(i)j}$ .
By construction, $T$ is a linear $G$-isometry that extends $t$.

	Now we prove that the two stated conditions are necessary.
	
	Case 1: Suppose that $G$ does not satisfy the UDP.  Consider  the sets $S, S'\subset V_i'$ and the linear codes  $\mathcal{C}_1 = span \{x\}$ and $\mathcal{C}_2 = span \{y\}$ as defined in the proof of Theorem \ref{thm:mac_1-level}. By construction, there is a linear map $t:\mathcal{C}_1 \rightarrow\mathcal{C}_2$ that preserves the $G$-weight. Suppose that  $t$ can be extended to $T\in GL(n, G)_{2}$. Then, Lemma \ref{lema_decomposicao} ensures that  $T$ defines a $G$-automorphism $\phi_{G}$ such that $\pi(\phi_{G}(\pi^{-1}(S))) = S'$, an absurd, since $\Gamma(S)$ and $\Gamma(S')$ are assumed to be non-isomorphic $L$-weighted subgraphs of $G'$.	
	
	Case 2: Suppose there are different elements $v_1,v_2,v_3\in V'$ such that $L(v_1)=L(v_2)=L(v_3)= l>1$. For $i=1,2,3$, let $e_{a_i}$ and $ e_{b_i}$ be two different vetors in $  \mathbb{F}_2^n$  such that $\pi(\mathrm{supp}(e_{v_i})) = \pi(\mathrm{supp}(e_{b_i}))=v_i$, which existence is ensured by the fact that $L(v_i)\geq 2 $. Let us define the codes $\mathcal{C}_1= span\{e_{av_1}+ e_{av_2}, e_{av_1}+ e_{av_3}, \}, \mathcal{C}_2= span\{e_{av_1}+ e_{av_3}, e_{bv_1}+ e_{bv_3}, \}\subset \mathbb{F}_2^n$. Since all the codewords has the same weight $2l$, any linear isomorphism $t$ between the codes  preserves the $G$-weight. However,  $3=|I_l^1|$ and $|I_l^2|=2$ so $t$ can not be extended by an isometry $T\in GL(n,G)_2$.
\end{IEEEproof}
\begin{theorem}
	Let $G(V,E)$ be a graph and suppose that $P_G$, the poset determined by  its reduced canonical form,  is a hierarchical poset.  Hence, $(\mathbb{F}_2^n,d_G)$ satisfies the MacWilliams Extension Property if, and only if, $G$ satisfies the UDP and the graph $\Gamma(V_i)$ satisfies the Condition $\Omega$, for each $1\leq i \leq h_G$.
\end{theorem}
\begin{IEEEproof}
	The canonical decomposition (Theorem \ref{thm:canonical_decomposition}) ensures  we may assume that $\mathcal{C}=\mathcal{C}_1 \oplus \cdots \oplus \mathcal{C}_{h_G}$ and $\mathcal{C}'= \mathcal{C}_1' \oplus \cdots \oplus \mathcal{C}_{h_G}'$.For each $c\in\mathcal{C}_i$, we have $\displaystyle t(c)=t_i(c) + F_i(c)$, where  $F_i: \mathcal{C}_i \rightarrow \sum_{j<i}\mathcal{C}_j'$ and $t_i:\mathcal{C}_i \rightarrow \mathcal{C}_i'$ are both linear maps, and $w_G(c)=w_G(t_i(c))$. Then, it is easy to verify that $t_i: \mathcal{C}_i\rightarrow\mathcal{C}_i'$ is also a linear isometry.
	Since $\mathrm{supp}(\mathcal{C}_i)$, $\mathrm{supp}(\mathcal{C}_i')\subset V_i$, we can consider   $\mathcal{C}_i,\mathcal{C}'_i\subset\mathbb{F}_q^{n_i}$ to be equipped with the metric based on graph $\Gamma(V_i)$ on $\mathbb{F}_q^{n_i}$, where $\Gamma(V_i)$ is the graph induced by $G$ on $V_i$. The previous proposition ensures that each $t_i$ admits an extension $T_i$ to $\mathbb{F}_q^{n_i}$, it means that, there is a linear $V_i$-isometry $T_i\in GL(n_i, \Gamma(V_i))_2$ of $\mathbb{F}_q^{n_i}$ into itself and  $\displaystyle T_i\vert_{\mathcal{C}_i} = t_i$.
		Since every linear map between linear codes can be extended to the entire space, it follows that the linear map $	T (	x_1+ \cdots + x_{h_G}) = (T_1+F_1)(x_1) + \cdots +(T_{h_G}+F_{h_G})(x_{h_G}) $ is a  $G$-isometry. 
		
		 Since we are assuming that $G'$ is hierarchical,  the UDP and the Condition $\Omega$ are checked at each level of the graph. If both the conditions holds up to the level $V_{j-1}$ of $G$ but one of them fails on level $V_j$, it is possible to construct codes $\mC_1$ and $\mC_2$ as in Proposition \ref{1level}, just taking the care to consider the generator vectors $x,y$ (in case the UDP does not hold) and the vectors $e_{ai},e_{bi}$ (in case condition $\Omega$ does not hold) to have the support on the $j$-level.
\end{IEEEproof}


\section*{Acknowledgment}
This work was supported in part by the S\~{a}o Paulo Research Foundation (FAPESP)
under grants no. 2014/10745-6, 2013/25977-7 and 2015/11286-8.


\end{document}